\documentclass[a4paper,fleqn,usenatbib]{mnras}

\usepackage{newtxtext,newtxmath}

\usepackage[T1]{fontenc}
\usepackage{ae,aecompl}

\usepackage{graphicx}	
\usepackage{amsmath}	
\usepackage{amssymb}	

\graphicspath{{./}{./plots/}}
\defcitealias{source:osterbrock2006}{OF06}
\defcitealias{source:dors2015}{D15}
\newcommand{\hb}{H$\beta$}
\newcommand{\ha}{H$\alpha$}
\newcommand{\llambda}{$\lambda\lambda$}

\title[Active Galaxy Abundances]{Chemical Abundances in Active Galaxies}

\author[Flury \& Moran]{
Sophia R. Flury$^{1}$\thanks{E-mail: sflury@umass.edu (SRF)}
and Edward C. Moran$^{2}$
\\
$^{1}$Department of Astronomy, University of Massechusetts Amherst, Amherst, MA 01002, United States\\
$^{2}$Department of Astronomy, Wesleyan University, Middletown, CT 06459, United States
}

\date{Accepted 2020 May 29. Received 2020 May 20; in original form 2019 November 12}

\pubyear{2020}

\begin{document}
\label{firstpage}
\pagerange{\pageref{firstpage}--\pageref{lastpage}}
\maketitle

\begin{abstract}
The Sloan Digital Sky Survey (SDSS) has proved to be a powerful resource for understanding the physical properties and chemical composition of star-forming galaxies in the local universe. The SDSS population of active galactic nuclei (AGN) remains as of yet less explored in this capacity. To extend the rigorous study of \ion{H}{ii} regions in the SDSS to AGN, we adapt methods for computing direct-method chemical abundances for application to the narrow-line regions (NLR) of AGN. By accounting for triply-ionized oxygen, we are able to more completely estimate the total oxygen abundance. We find a strong correlation between electron temperature and oxygen abundance due to collisional cooling by metals. Furthermore, we find that nitrogen and oxygen abundances in AGN are strongly correlated. From the metal-temperature relation and the coupling of nitrogen and oxygen abundances, we develop a new, empirically and physically motivated method for determining chemical abundances from the strong emission lines commonly employed in flux-ratio diagnostic diagrams (BPT diagrams).  Our approach, which for AGN reduces to a single equation based on the BPT line ratios, consistently recovers direct-method abundances over a 1.5 dex range in oxygen abundance with an rms uncertainty of 0.18 dex.  We have determined metallicities for thousands of AGN in the SDSS, and in the process have discovered an ionization-related discriminator for Seyfert and LINER galaxies.
\end{abstract}

\begin{keywords} 
ISM: abundances -- galaxies: abundances -- galaxies: active
\vspace{10pt}
\end{keywords}



\section{Introduction}

Spurred by revolutionary ideas about the nature of gaseous nebulae \citep[e.g.,][]{source:stromgren1939}, Menzel applied then-new ideas about quantum mechanics to emission-line phenomena to develop a working theory of nebular astrophysics, often in collaboration with Baker and Aller. This enabled the first empirical estimates of physical properties like electron temperature \citep{source:menzeletal1941} and ionic abundances like O$^{+2}$/H \citep{source:menzelaller1941} in gaseous nebulae. Seyfert, their contemporary, reported the discovery of a new class of emission-line objects in 1943 that later came to be known as active galactic nuclei (AGN). While Menzel's theory, amended and expanded by astronomers over the following decade or so, has grown commonplace in the research of \ion{H}{ii} regions, there has been comparatively little application to AGN, preventing the sort of large population studies enjoyed by their line-emitting cousins.

This disparity is perhaps understandable for two reasons:\ (i) \ion{H}{ii} regions far outnumber AGN, and many are easily studied due to their location within the Milky Way and other Local Group galaxies; and (ii) AGN nebulae have a more complex ionization structure \citep[e.g.,][henceforth OF06]{source:osterbrock2006}. Sixteen years passed between the development of a method for the direct estimation of chemical abundances in the Orion Nebula \citep{source:aller1959} and the first investigation of metallicity in an AGN \citep{source:osterbrock1975}. While \cite{source:koski1978} and \cite{source:shuderoster1981} analyzed the physical properties of the gas in Seyfert nebulae, and \cite{source:storchiberg1990} established the range of metallicities for a sample of AGN using photoionization models, it was not until \cite{source:cruzgonz1991} that direct-method abundances (i.e., abundances determined using the electron temperature and multi-level atomic models) were computed for a population of Seyferts, a full 50 years after the introduction of the technique.

Because the ionization structure of AGN is complex and highly ionized species are present, high-quality observations and a detailed physical picture of  photoionized nebulae in AGN are required for accurate estimates of their elemental abundances. As a possible workaround, it has been demonstrated that the metallicities of the narrow-line regions (NLRs) of Seyfert nuclei can be inferred from the chemical abundances of nuclear starbursts within the same host galaxies \citep[e.g.,][]{source:evansdop1987,source:storchiberg1998}. Similarly, AGN abundances appear to correlate with radial metallicity gradients observed in the extranuclear \ion{H}{ii} regions of some Seyfert galaxies \citep{source:dors2015}.  However, not all AGN host galaxies have nuclear or extranuclear star-forming regions from which nuclear metallicities can be gleaned --- indeed, many are gas-poor ``red-and-dead'' objects such as bulge-dominated lenticular galaxies. Furthermore, the weakness of the temperature-sensitive [\ion{O}{iii}]~$\lambda$4363 auroral line can impede efforts to constrain the electron temperature and, therefore, direct-method abundances.

To circumvent these obstacles, astronomers have turned to radiative transfer codes to model the physical and chemical properties of nebulae based on  observed emission-line fluxes. This method takes one of several possible forms. The first involves assembling a large set of AGN photoionization models to establish trends between metallicity and line flux ratios so that abundances can be inferred from readily observable emission lines (the ``strong'' emission lines, or SELs) for an arbitrary AGN sample \citep[e.g.,][]{source:storchiberg1998,source:groves2004a,source:perezmontero2019}. The second approach requires detailed photoionization modeling of individual AGN, such as the NLR in the Circinus galaxy \citep{source:oliva1998} or the literature sample of \cite{source:dors2017}. This approach allows a number of parameters to be varied, including the abundances of individual elements, in order to best predict the observed SEL fluxes.  However, while robust in treating the relevant physics, radiative transfer codes such as \textsc{Cloudy} \citep{source:cloudy2013,source:cloudy2017} and \textsc{Mappings} \citep{source:mappings2013,source:mappings2017} rely on a significant number of assumptions about the nebular environment. Unfortunately, the uncertanties associated with some of the necessary atomic data are substantial, which can introduce systemic error into the emission-line flux predictions. Likewise, information about the shape of the ionizing continuum, the geometry and thermodynamic structure of the nebula, the properties of the dust present, and the equilibrium conditions are required, all of which can influence the predicted strong-line fluxes. Such information is frequently unavailable from observations and must, therefore, be assumed in order to obtain chemical abundances by this approach.

The goal of this study is to develop a robust technique for determining direct-method chemical abundances based on the strong, commonly available emission lines that are measured in large spectroscopic surveys. A reliable direct metallicity method will allow statistically rigorous investigations of the physical and chemical environments of AGN, which ultimately will provide more general insight into the enrichment histories of galactic nuclei.  We begin by defining samples of AGN and star-forming galaxies appropriate for such work in \S~\ref{s:samples}. In \S~\ref{s:phys}, we present the physical characteristics of the objects in the these samples, which, in conjunction with a multiple ionization-zone model, are used in \S~\ref{s:abn} to estimate total relative chemical abundances.  From these results, we then investigate the effect of cooling by metals on the electron temperature in \S~\ref{s:cooling} and the scaling of nitrogen with oxygen in \S~\ref{s:nitrogen}. In \S~\ref{s:selmethod}, we present a simple model for common strong emission line flux rations from which we develop a new technique for determining chemical abundances in AGN. We verify this method with a comparison between our new method and the direct-method abundances. We then apply this technique to a large sample of AGN and discuss the implications.

\section{Sample Definition}\label{s:samples}

In order to perform population analyses of any import, we need a sample of galaxies whose nuclear activity is reliably classified according to ``BPT'' emission-line flux-ratio diagrams \citep{source:bpt1981,source:vo1987}. Therefore, we must select objects for which the emission lines necessary for nebular diagnostics are detected in the optical spectrum.

To begin, we employ published emission-line fluxes of galaxies from the Portsmouth processing \citep{source:portsmouth2013} of the eighth data release (DR8) of the Sloan Digital Sky Survey \citep[SDSS;][]{source:sdssdr8}. In their spectral analysis, the starlight continuum was modeled using penalized pixel fitting \citep[\textsc{ppxf};][]{source:ppxf2004} as implemented by the \textsc{Gandalf} code \citep{source:gandalf2006} with \cite{source:maraston2011} synthetic stellar population (ssp) templates and the \cite{source:calzetti2001} extinction law. Emission lines were simultaneously fitted with Gaussian profiles to obtain their fluxes (which we corrected for reddening following the method described in Appendix A). Using these measurements, we have defined a sample of DR8 objects that enable us to investigate the chemical abundances in AGN.

\subsection{Parent Samples}\label{s:parentsample}

\begin{figure*}
	\includegraphics[width=\linewidth]{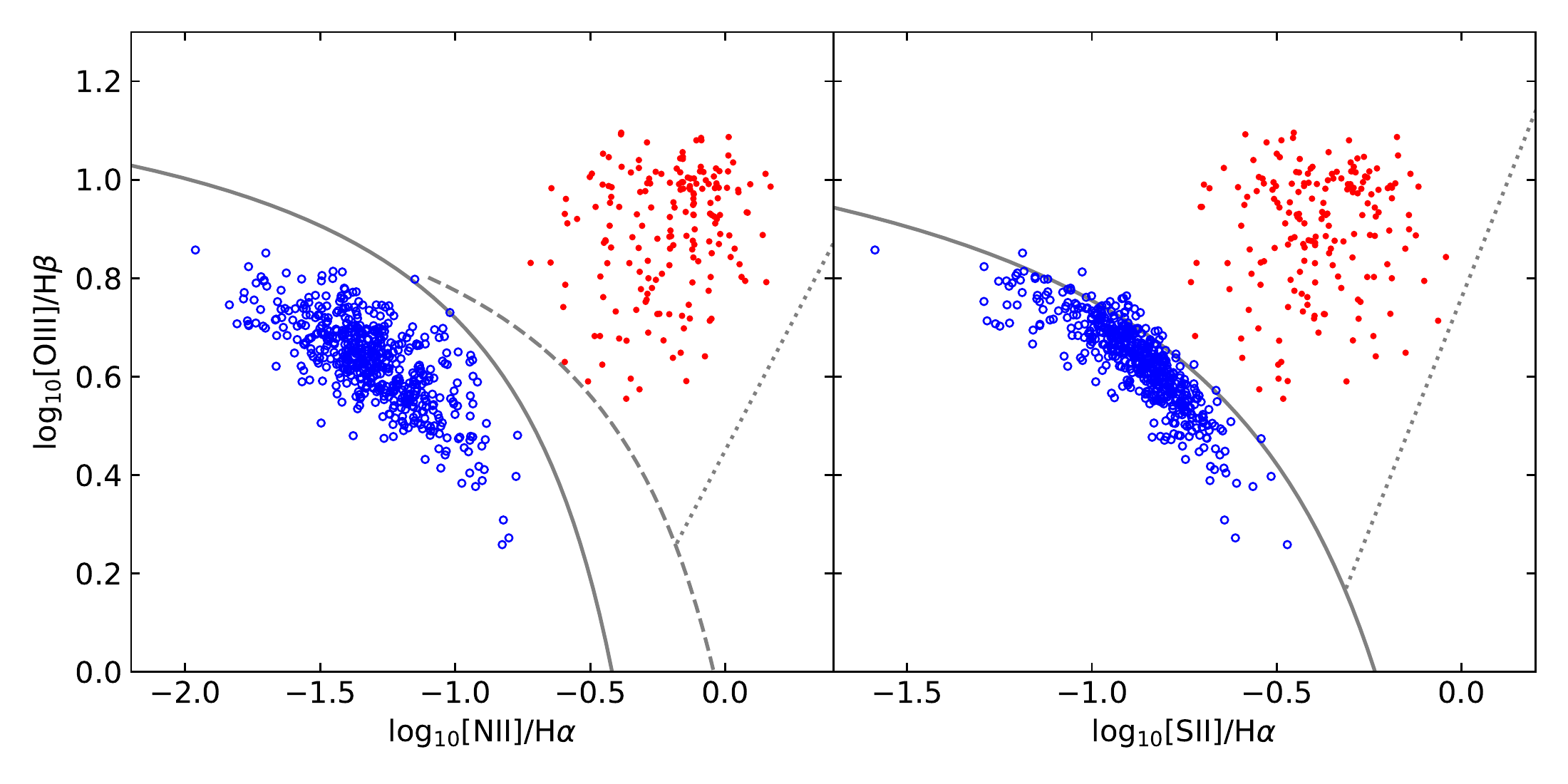}
    \caption{BPT diagnostic diagrams for our samples of AGN (filled red symbols) and \ion{H}{ii} galaxies 
 (open blue symbols) from the SDSS DR8. The solid curves indicate the empirical demarcations between 
 \ion{H}{ii} galaxies and AGN defined by \citet{source:kauffmann2003} and \citet{source:kewley2006}. 
 The dashed curve on the [\ion{N}{ii}]/\ha\ plot represents the theoretical ``maximum starburst'' line 
 from \citet{source:kewley2001}. In the AGN zones, empirical divisions between Seyfert galaxies and 
 LINERs \citep{source:schawinski2007,source:kewley2006} are indicated by dotted lines.}
    \label{fig:bpt}
\end{figure*}

Our parent sample is based on the detection of six prominent SELs: \hb\ $\lambda 4861$, [\ion{O}{iii}]~$\lambda5007$, \ha\ $\lambda 6563$, [\ion{N}{ii}]~$\lambda$6583, and [\ion{S}{ii}]~\llambda6716,31. We require all of these lines to be detected with an amplitude-over-noise (A/N) value of at least 3. This allows us to classify the nuclear activity of the objects using the [\ion{N}{ii}]/\ha\ vs.\ [\ion{O}{iii}]/\hb\ and [\ion{S}{ii}]/\ha\ vs.\ [\ion{O}{iii}]/\hb\ BPT diagnostic diagrams (see Fig.\ \ref{fig:bpt}). Our classification criteria are as follows. Star-forming galaxies are defined to be objects whose flux ratios fall below and to the left of the empirical curves that demarcate \ion{H}{ii} and AGN zones on the BPT diagrams \citep{source:kauffmann2003,source:kewley2006}. Objects that
are no more than 0.05 dex normal-to-curve over the \cite{source:kewley2006} line on the [\ion{S}{ii}]/\ha\ BPT diagram are also considered to be \ion{H}{ii} galaxies.  AGN are defined to be objects with line ratios that place them above and to the right of the theoretical ``maximum starburst'' line on the [\ion{N}{ii}]/\ha\ plot \citep{source:kewley2001} and at least 0.05 dex above the corresponding demarcation on the [\ion{S}{ii}]/\ha\ diagram.  

For the vast majority of objects, the classifications according to both of the BPT diagnostics are consistent; these make up our ``\ion{H}{ii}'' (152,322 objects) and ``AGN'' (8,720 objects) parent samples.

The redshifts of the galaxies in the parent sample have a maximum value of 0.370 with a median of 0.075.  Assuming $H_0=73$ km s$^{-1}$ Mpc$^{-1}$, the bulk of the objects in the sample lie at distances between 180 Mpc and 420 Mpc.

\subsection{Chemical Abundances Samples}\label{s:subsample}

From the parent sample, we culled objects based on criteria for several emission lines to assemble \ion{H}{ii} and AGN subsamples that can be used to measure chemical abundances. First, we impose a detection of the (weak) [\ion{O}{iii}]~$\lambda4363$ auroral line, which is necessary for the determination of properties such as electron temperature ($T_e$) and density ($n_e$) of a photoionized nebula.  We further require that [\ion{O}{iii}]~\llambda4959,5007/[\ion{O}{iii}]~$\lambda$4363 $> 16$ for the galaxies.  This eliminates a small number of objects with improbably high electron temperature estimates ($T_e > 4\times10^4$ K for $n_e\sim10^3$ cm$^{-3}$) due to errors in their reported [\ion{O}{iii}]~$\lambda$4363 fluxes.  We also impose the condition that $\sigma_{T_e}/T_e < 1$, where $\sigma_{T_e}$ is the uncertainty in the derived electron temperature (see \S~\ref{s:phys}). This is to ensure that the values of $T_e$ and $n_e$ we obtain are reasonably well constrained.

We note that these requirements introduce an electron temperature bias for this subset of objects.  The relative strength of [\ion{O}{iii}]~$\lambda$4363 increases sharply at $T_e \gtrsim 10^4$ K \citepalias[e.g.,][]{source:osterbrock2006}, so the auroral line tends to be more easily detected in high-$T_e$ objects. As discussed by \citet{source:izotov2006}, a consequence of this bias is that the \ion{H}{ii} galaxies selected occupy only the ``upper'' portion of the star-forming locus on BPT diagnostic diagrams (where [\ion{O}{iii}]/\hb\ $\gtrsim 2$).  This is the lower metallicity portion of the \ion{H}{ii} distribution, so the correlations we investigate below do not extend to the highest expected metal abundances in star-forming regions.

From the galaxies with detected [\ion{O}{iii}]~$\lambda$4363, we have selected objects with a detected [\ion{O}{ii}]~\llambda3726,29 doublet, which is needed to compute oxygen abundances via the direct method. We further limit the sample to those objects for which the the derived oxygen abundance (discussed in \S~4.2) is fairly well constrained, i.e., $\sigma_{\text{[O/H]}} <$ 0.5~dex.  The final \ion{H}{ii} and AGN samples for abundance measurements contain 539 and 180 galaxies, respectively.  Due to the blue wavelength cutoff of $\sim 3800$ \AA\ in SDSS spectra, all of the objects in these samples have redshifts above $z \approx 0.02$. However, their SEL flux ratios are evenly distributed with respect to those displayed by the parent sample, suggesting that the redshift limit does not bias our abundance investigation in any way. 

We note that all objects classified as AGN in this subsample have [\ion{O}{iii}]/\hb\ $> 3$ \citep[the traditional threshold for Seyfert galaxies;][]{source:vo1987} and fall in the ``Seyfert'' zones on BPT diagrams defined by \cite{source:kewley2006} and \cite{source:schawinski2007}.  Thus, we can be confident that these objects are powered by accretion onto supermassive black holes.

\section{Physical Properties}\label{s:phys}

Our assessment of the nebular conditions in the \ion{H}{ii} and AGN samples employs the five-level atom model for computing the electron density and temperature. For each object, we begin by solving the equilibrium equations for the bound electronic states that correspond to the observed radiative transitions using the open-source software \textsc{PyNeb} \citep{source:pyneb}, then calculate $n_e$ and $T_e$ iteratively using the measured extinction-corrected [\ion{S}{ii}] and [\ion{O}{iii}] line fluxes until the results converge on a consistent set of solutions.

\subsection{Electron Density}\label{s:edens}

Assuming collisional equilibrium, application of the five-level atom model allows electron density to be computed from the ratio of the fluxes associated with the $^2$D$^0_{5/2}\to^4$S$^0_{3/2}$ and $^2$D$^0_{3/2}\to^4$S$^0_{3/2}$ radiative transitions. The [\ion{S}{ii}]~\llambda6716,31 and [\ion{O}{ii}]~\llambda3726,29 doublets are the common optical emission lines employed to obtain electron density. Because the [\ion{O}{ii}] doublet is unresolved in the low-resolution SDSS spectra, we derive electron densities using the [\ion{S}{ii}] doublet.
From the equation for collisional equilibrium \citepalias[see, e.g., Eq.~3.23 in][]{source:osterbrock2006}, the ratio of the [\ion{S}{ii}] doublet emission-line fluxes has the form
\begin{equation}\label{eq:edens}
\frac{j_{\lambda6716}}{j_{\lambda6731}} = a_n\frac{1+b_n~n_e}{1+c_n~n_e}
\end{equation}
where $a_n$ depends on the ratio of collision strengths\footnote{Following discussion in Ch.~5.6 of \citetalias{source:osterbrock2006}, $a_n$ is determined primarily by the ratio of the $^2$D$^0$ states' statistical weights $g=2(s+\ell)+1$. For the [\ion{S}{ii}] doublet, $a_n\approx 3/2$.} and $b$ and $c$ depend on the ratio of state population to state de-population rates. Values of these parameters for the relevant range of electron temperatures are listed in Table \ref{tab:s2params}.

\begin{table}
	\centering
	\caption{Parameters for [\ion{S}{ii}] doublet ratio vs. electron density for a range of electron temperatures.}
	\label{tab:s2params}
	\begin{tabular}{cccc} 
		\hline
		$T_e$ ($10^4$ K) & $a_n$ & $b_n$  (cm$^3$)& $c_n$ (cm$^3$)\\
		\hline
		0.5 & 1.471 & 6.710 & 2.259 \\
		1.0 & 1.450 & 4.725 & 1.591 \\
		1.5 & 1.431 & 3.854 & 1.298 \\
		2.0 & 1.417 & 3.326 & 1.122 \\
		2.5 & 1.406 & 2.967 & 1.001 \\
		3.0 & 1.396 & 2.700 & 0.909 \\
		3.5 & 1.391 & 2.483 & 0.837 \\
		\hline
	\end{tabular}
\end{table}

We compute the electron density for a given electron temperature by root-finding the value of $n_e$ that minimizes the difference between the observed doublet flux ratio and that predicted by \textsc{PyNeb}. Assuming $T_e=10^4$ K, this method provides an initial estimate of electron density from the [\ion{S}{ii}] flux ratio that will need to be recomputed iteratively in tandem with determinations of electron temperature.

\subsection{Electron Temperature}\label{s:etemp}

For a photoionized gas in collisional equilibrium, the five-level atom model allows electron temperature to be computed from the ratio of the fluxes associated with the
$^1$D$_{2}\to^3$P and $^1$S$_{0}\to^1$D$_{2}$ radiative transitions.  The
[\ion{O}{iii}]~\llambda4959,5007 doublet and [\ion{O}{iii}]~$\lambda4363$ auroral line are typically used to derive the electron temperature because [\ion{O}{iii}]~$\lambda4363$ is often the only diagnostic auroral emission line detected in a spectrum.

\begin{table}
	\centering
	\caption{Parameters for [\ion{O}{iii}] doublet-auroral line ratio vs. electron temperature for a range of electron densities.}
	\label{tab:o3params}
	\begin{tabular}{cccc} 
		\hline
		$n_e$ (cm$^{-3}$) & $a_T$ & $b_T\times10^4$ K & $c_T$ (K$^{1/2}$~)\\
		\hline
		10         & 7.995 & 3.331 & ~8.937 \\
		$10^2$ & 7.987 & 3.330 & ~8.844 \\
		$10^3$ & 7.964 & 3.330 & ~8.970 \\
		$10^4$ & 7.995 & 3.341 & 14.913 \\
		\hline
	\end{tabular}
\end{table}

As with the electron density, we use \textsc{PyNeb} to compute the [\ion{O}{iii}] flux ratio for a given $T_e$ and assumed $n_e$ and calculate the temperature which minimizes the difference between the predicted and observed flux ratio. For reference, we express the [\ion{O}{iii}] temperature diagnostic following Eq.~5.4 in \citetalias{source:osterbrock2006}:
\begin{equation}\label{eq:etemp}
\frac{j_{\lambda4959}+j_{\lambda5007}}{j_{\lambda4363}} =
\frac{a_T~ \exp\left(b_T~ T_e^{-1}\right)}
{1+ c_T~ T_e^{-1/2}}
\end{equation}
where $a_T$ depends primarily on the ratio of ($^1$D$_{2}$,$^3$P) collision strengths and Einstein coefficients for spontaneous radiation to those of ($^1$S$_{0}$,$^1$D$_{2}$), $b_T$ scales the electron temperature by the energy difference between the $^1$S$_{0}$ and $^1$D$_{2}$ states to account for collisional excitation to $^1$S$_{0}$, and $c_T$ contains a density term to account for collisional de-excitations.
Values of these parameters for a range of electron densities are listed in Table \ref{tab:o3params}.

\subsection{Temperature-Density Coupling}\label{s:tempdens}

As suggested by the above discussion, the electron temperature must be known or assumed to determine the electron density, and vice versa.  To address this, we have developed an iterative numerical procedure for computing $n_e$ in tandem with $T_e$, which yields stable values of both quantities that are consistent with the observed emission-line flux ratios.  While \textsc{PyNeb} includes a method for simultaneously solving for $n_e$ and $T_e$ for a given set of flux ratios, it does not account for differences in temperature between the [\ion{S}{ii}]- and [\ion{O}{iii}]-emitting zones in a nebula. Our approach, while similar, accounts for these temperature differences.

We begin by computing the electron temperature using a bisection routine, which finds the root corresponding to the difference between the observed [\ion{O}{iii}]~\llambda4959,5007/$\lambda$4363 flux ratio and that predicted by \textsc{PyNeb} for the low-density limit (where $n_e \to 0$ cm$^{-3}$). Then, using a second bisection routine, we compute the electron density by finding the root of the difference between the [\ion{S}{ii}]~\llambda6716,31 flux ratio and that predicted by \textsc{PyNeb}. For this calculation, the [\ion{O}{iii}]-based estimate of $T_e$ is used as input, but it must first be scaled to a value appropriate for the zone where O$^{+}$ and S$^{+}$ ions reside using the semi-empirical relation
\begin{equation}\label{eq:lowTe}
T_{e,{\rm[O~II]}} = \left(\frac{6930}{T_{e,{\rm[O~III]}}} + 2810\,\,\text{K}^{-1}\right)^{-1}
\end{equation}
from \cite{source:perezmontero2003}. We then re-compute the electron temperature using the revised estimate of the electron density, and so on. Provided the [\ion{S}{ii}] flux ratio is theoretically permitted (i.e., between $\sim 0.4$ and $\sim 1.5$), this procedure always converges on stable solutions, typically within three to five iterations depending on the tolerance. [\ion{S}{ii}] flux ratios which fall outside the theoretically permitted range occur in 126 \ion{H}{ii} galaxies and 8 AGN in our sample. For these objects, we assign a density of 100 cm$^{-3}$ following precedent \citep[e.g.,][]{source:vanzee1998a,source:dors2013} as this value is plausible for either class of objects under investigation.  

\begin{figure}
	\includegraphics[width=\columnwidth]{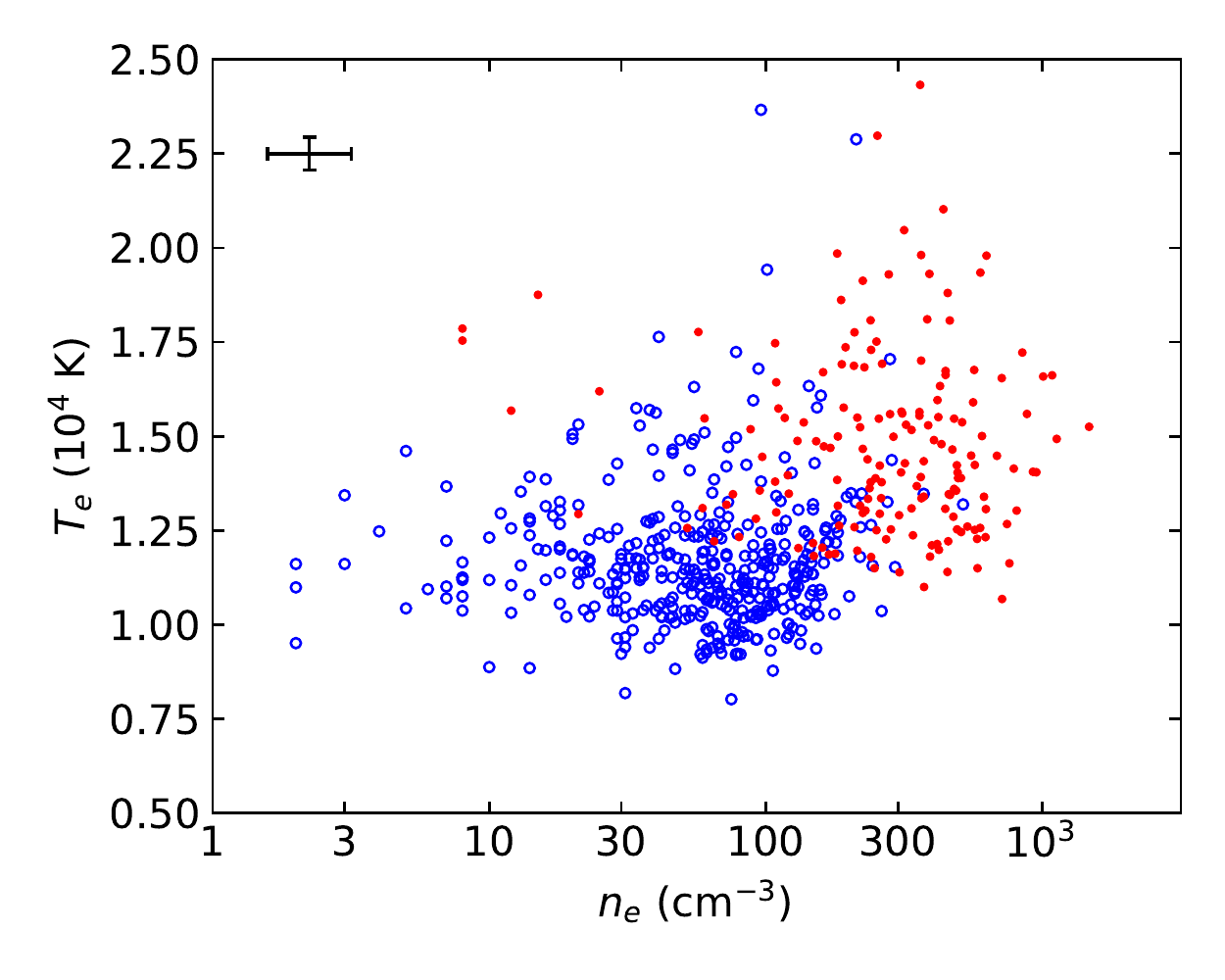}
    \caption{Electron temperatures and densities for the \ion{H}{ii} galaxies (open blue circles) and AGN (filled red circles) in our sample of SDSS objects.  Typical 1~$\sigma$ uncertainties for both quantities are shown in the upper left corner.}
    \label{fig:temden}
\end{figure}

Electron temperatures and densities for the galaxies in our sample are plotted in Figure \ref{fig:temden}. Although temperature and density are formally coupled due to the physics of equilibrium, the two properties are not correlated for either \ion{H}{ii} galaxies or AGN. However, it is clear that the NLR environments of AGN tend to be hotter and denser than \ion{H}{ii} regions near the centers of star-forming galaxies. We will return to this point in \S5. Excluding objects with [\ion{S}{ii}] flux ratios that fall outside the theoretical limits, the median density and temperature (with associated dispersions) for the \ion{H}{ii} galaxies are $\log(n_e/{\rm cm}^{-3}) = 1.924 \pm 0.399$ and $T_e = (1.15 \pm 0.23) \times 10^4$~K.
For AGN, we find $\log(n_e/{\rm cm}^{-3}) = 2.456 \pm 0.382$ and $T_e = (1.43 \pm 0.34) \times 10^4$~K.

\section{Gas Phase Abundances}\label{s:abn}

To compute gas phase abundances, we employ the direct method using the electron temperatures and densities determined above. For a complete estimate of the relative oxygen abundance, we adopt a three-ionization-zone approximation for nebulae in which oxygen exists in singly, doubly, and triply ionized species. We measure ionic abundances for singly and doubly ionized oxygen and account for triply ionized oxygen with an ionization correction factor. Finally, we validate our approach by comparing our results against those reported for a literature sample of AGN.

\subsection{Justification for the Three-Zone Model}\label{s:threezone}

The relative amount of O$^{+3}$ in \ion{H}{ii} regions is usually negligible \citep[e.g.,][]{source:peimbercost1969,source:kobulnicky1999,source:kennicutt2003} except in extreme starbursts with a sizable Wolf-Rayet population \citep[e.g.,][]{source:izotov2006}. However, studies of planetary nebulae, which have harder ionizing spectra than OB associations, have found that O$^{+3}$, while not the most abundant species of the element, consists of a non-negligible fraction of the total oxygen with an upper estimate of 20\% in extreme cases \citep[e.g.,][]{source:burgess1960,source:rubin1997}.

This must certainly be the case in a Seyfert NLR:\  the ionizing continuum produced by the central engine possesses significantly more energy per photon than that emitted by OB associations.  Indeed, several studies of the [\ion{O}{iv}] $\lambda$25.89$\mu$m mid-infrared emission line have found the line to be far stronger in AGN (particularly Seyferts) than in starburst nuclei \citep{source:diamstan2012,source:fernonti2016} and that the luminosity of [\ion{O}{iv}] correlates strongly with X-ray and infrared luminosities \citep{source:garcbern2016}.
Given the significance of the [\ion{O}{iv}] line in AGN, it stands to reason that the summed ionic abundances must account for O$^{+3}$ in order to appropriately estimate the total oxygen abundance in AGN. Because of the extended partially-ionized zones in AGN, as much oxygen might be contained in the O$^{0}$ as in O$^{+3}$ zone. However, hydrogen and neutral oxygen have very similar ionization potentials (13.598 eV and 13.618 eV, respectively). These neutral species are further coupled through charge-exchange reactions, meaning that
\begin{equation}
\frac{\text{O}}{\text{H}} = \left(\frac{\text{O}^{0}}{\text{H}^{0}}\right)
\end{equation}
is necessarily true \citep[cf.][]{source:peimbercost1969,source:kobulnicky1999}. However, as there is no means to estimate H$^0$ from an optical spectrum alone, measurements of the [OI]$\lambda6300$ flux cannot be used to infer the chemical abundance. As such, only the ionized oxygen is included in estimates of the oxygen abundance relative to H$^{+}$. This gives a final relation for the total relative oxygen abundances of
\begin{equation}\label{eq:3zone}
\frac{\text{O}}{\text{H}} = \frac{\text{O}^{+}+\text{O}^{+2}+\text{O}^{+3}}{\text{H}^{+}}.
\end{equation}
With this established, the remaining work in determining the total oxygen abundance lies in deriving values for the O$^{+}$, O$^{+2}$, and H$^{+}$ ionic abundances and, subsequently, inferring that of O$^{+3}$ with an ionization correction factor (ICF).

\subsection{Direct Method Ionic Abundances}\label{s:dmabn}

While some studies find discrepancies between the direct method for computing abundances from collisionally excited lines (CELs) and recombination lines \citep[e.g.,][]{source:esteban2009,source:torbsancip2017} and photoionization models \citep[e.g.,][]{source:kewlell2008,source:dors2015}, an appropriate treatment of ionization structure can account for at least some of the disagreement in the case of Seyferts. Photoionization models rely on assumptions about, among other things, the shape of the ionizing spectrum, dust depletion rates, boundary conditions, and initial conditions, and are also error prone in cases where photoionization cross-sections and collisional strengths are uncertain. Recombination lines are less sensitive to temperature and thermal inhomogeneities than CELs but require exceptionally high SNR spectroscopy to be measured with any confidence.

The direct method of determining abundances from CELs relies on the emissivity
\begin{equation}\label{eq:colemit}
j_\lambda = n_{\chi^i} n_e \varepsilon_\lambda(n_e,T_e)\frac{4\pi\lambda}{hc},
\end{equation}
where $n_{\chi^{i}}$ is the abundance of a particular element $\chi$ in ionized species $i$, and $\varepsilon_\lambda$ is the emission coefficient \citepalias[Eq.\ 5.41 in][]{source:osterbrock2006}. We compute emission coefficient $\varepsilon_\lambda$ using \textsc{PyNeb}, which follows Equation 3.20 in \citetalias{source:osterbrock2006}. For recombination emission lines, the assumed relationship from Equation 4.14 in \citetalias{source:osterbrock2006} is
\begin{equation}\label{eq:recemit}
j_\lambda = n_{\chi^i} n_e \alpha_{eff}\frac{4\pi\lambda}{hc}
\end{equation}
where $\alpha_{eff}$ is the effective recombination rate. This effective rate accounts for collisions in the transitional cascade, which populates the state wherefrom an electronic transition occurs to produce a photon at wavelength $\lambda$. We use \textsc{PyNeb} to compute the Case B recombination emissivities for hydrogen.

Together with the physical properties estimated in \S~\ref{s:phys}, we compute $\varepsilon_\lambda$ for the [\ion{O}{iii}] \llambda4959,5007 doublet and $\alpha_{eff}$ for H$\beta$ to determine the relative abundance of doubly ionized oxygen, such that
\begin{equation}\label{eq:opphabund}    
\frac{\text{O}^{+2}}{\text{H}^+} = \frac{(j_{4959}+j_{5007})\alpha_{4861}}{(\varepsilon_{4959}+\varepsilon_{5007})j_{4861}}.
\end{equation}
Because we lack measurements of the auroral [\ion{O}{ii}] lines, we again use the semi-empirical relation from \cite{source:perezmontero2003} to estimate the [\ion{O}{ii}] electron temperature and compute the emission coefficients for the [\ion{O}{ii}]~\llambda3726,29 doublet (see Equation \ref{eq:lowTe} above). This allows us to determine the relative abundance of singly ionized oxygen:
\begin{equation}\label{eq:ophabund}
\frac{\text{O}^{+}}{\text{H}^+} = \frac{(j_{3726}+j_{3729})\alpha_{4861}}{(\varepsilon_{3726}+\varepsilon_{3729})j_{4861}}.
\end{equation}
Combined, these two ionic abundances give an initial lower limit to the total oxygen abundance but will need to be corrected to account for more highly ionized species.

We also compute the nitrogen abundance relative to oxygen using N$^{+}$ and O${+}$ ions. The N$^{+}$ and O${+}$ ions occupy similar ionization zones in a photoionized nebula, as indicated by the similarities in both neutral and singly-ionized species' ionization potentials, allowing these to be used as a proxy for the nitrogen abundance relative to oxygen \citep[e.g.][]{source:peimbercost1969}. We proceed with the assumption that N/O = N$^+$/O$^+$ \citep{source:peimbert1968,source:peimbercost1969}. From Equation 18 in \citet{source:peimbert1968}, we obtain
\begin{equation}
\frac{\text{N}}{\text{H}}  = \frac{\text{N}}{\text{O}}\times \frac{\text{O}}{\text{H}}.
\end{equation} 
As was done with the relative oxygen abundances, we compute collisional emission coefficients using \textsc{PyNeb} to determine the relative ionic abundance ratio
\begin{equation}\label{eq:noabund}
\frac{\text{N}^{+}}{\text{O}^+} = \frac{(j_{6548}+j_{6583})(\varepsilon_{3726}+\varepsilon_{3729})}
{(j_{3726}+j_{3729})(\varepsilon_{6548}+\varepsilon_{6583})}.
\end{equation}
Because the low ionization zones for the two elements largely occupy the same region of the nebula, we assume when computing the emissivities that the electron temperature for singly-ionized nitrogen is the same as that of singly-ionized oxygen.

\subsection{Ionization Correction Factor}\label{s:icf}

It is necessary, especially in the case of AGN, to account for oxygen locked up in unobserved ionic species like O$^{+3}$ using an ICF. The similarity between the O$^{+2}$ and He$^+$ ionization potentials (54.936 eV and 54.417 eV, respectively) allows one to write
\begin{equation}
\frac{\text{O}^{+3}}{\text{O}^{+}+\text{O}^{+2}+\text{O}^{+3}} 
\propto  \frac{\text{He}^{+2}}{\text{He}^{+}+\text{He}^{+2}}
\end{equation}
in order to infer the abundance of O$^{+3}$. Because He$^0$ ionizes at 24.59 eV while O$^0$ ionizes at 13.62 eV, a proportionality factor $p_{\text{ICF}}$ is necessary to complete the relation by accounting for the difference in ionization structure for the two elements. O$^{+}$ and O$^{+2}$ have an average ionization potential of 24.369 eV, which is close to that of neutral helium. It stands to reason, then, that taking the mean of the O$^{+}$ and O$^{+2}$ abundances can correct for the mismatch in the O and He ionization zones. This gives $p_{\text{ICF}}=0.5$, as suggested by \citet{source:izotov2006}, resulting in a final ICF of
\begin{equation}
\text{O}^{+3} = 0.5\frac{y}{1-y}\left(\text{O}^{+}+\text{O}^{+2}\right),
\end{equation}
where $y=\text{He}^{+2}(\text{He}^{+}+\text{He}^{+2})^{-1}$.

\begin{figure}
	\includegraphics[width=\columnwidth]{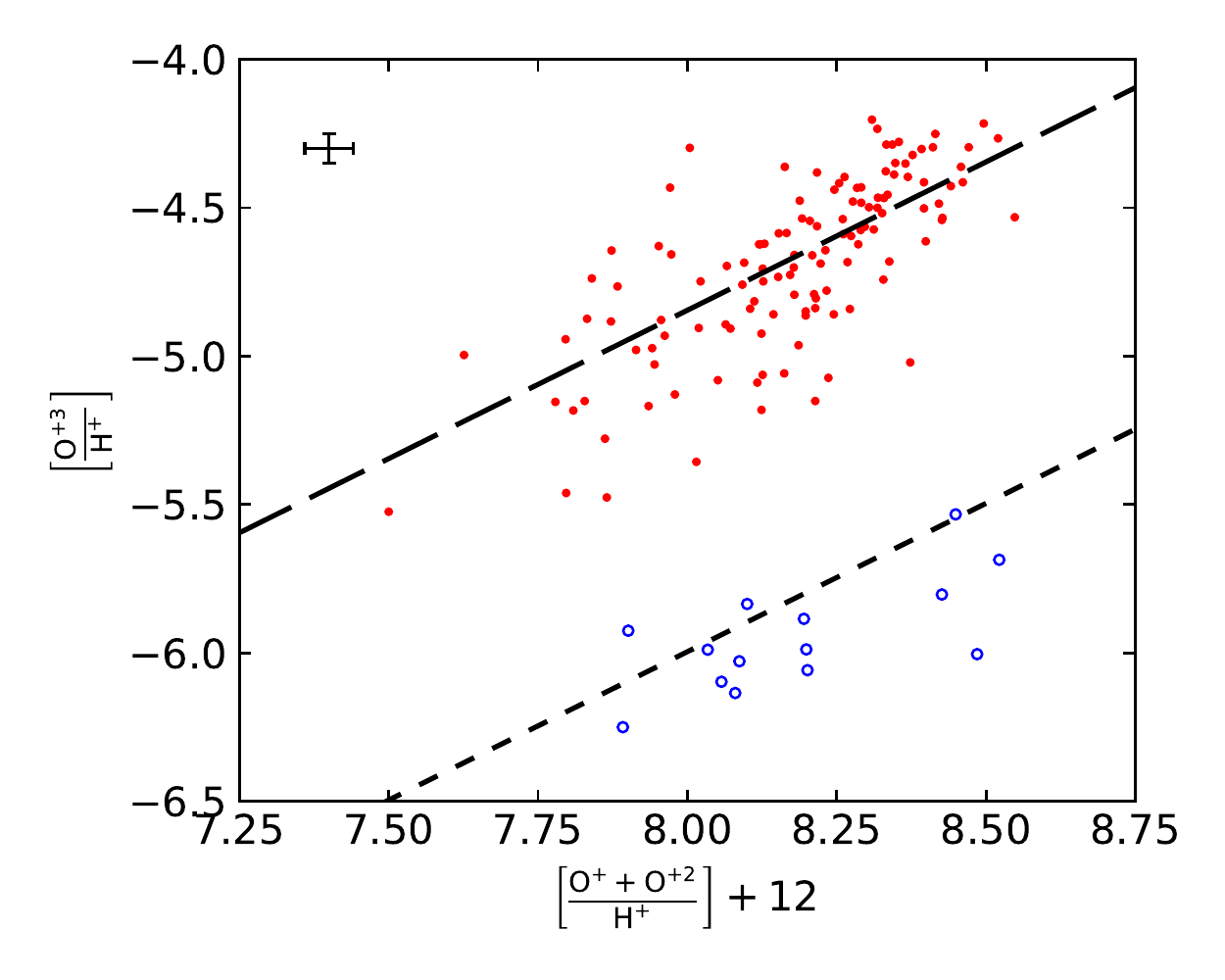}
    \caption{The [O$^{+3}$/H$^+$] abundance as a function of two-zone oxygen abundances for objects in our sample for which the fluxes of \ion{He}{1}~$\lambda$5875 and \ion{He}{2}~$\lambda$4686 are measured. The short-dashed line represents 1\% of the oxygen locked up in the triply ionized species, while the long-dashed line represents 12.5\%. Symbols are the same as in Figure \ref{fig:bpt}.}
    \label{fig:icf}
\end{figure}

We apply this ICF to all objects in our sample for which \ion{He}{1}~$\lambda$5875 and \ion{He}{2}~$\lambda$4686 fluxes have been measured (with A/N $>2$) by the Portsmouth group.  This comprises 125 AGN and 13 \ion{H}{ii} galaxies. For these objects, the [O$^{+3}$/H$^+$] ionic abundances estimated by this method are shown in Figure \ref{fig:icf}. As the Figure indicates, there is a clear difference in the typical [O$^{+3}$/H$^+$] values for Seyfert and star-forming galaxies. For the AGN, $\sim12.5$\% of the total ionic oxygen abundance exists in the O$^{+3}$/H, while for \ion{H}{ii} galaxies the percentage is $\sim1$\% at best.

For galaxies lacking helium line fluxes, we use the fractional O$^{+3}$ implied by the typical ICF measured for objects of the same activity class, and estimate He$^+$ and He$^{++}$ using the \ion{He}{1} and \ion{He}{2} Case B recombination coefficients computed by \textsc{PyNeb} with the O$^{+2}$ electron temperatures.  With values of [O$^{+3}$/H$^{+}$] for all objects in our sample, we now have abundances for each species necessary to solve for the oxygen abundance relative to hydrogen using Equation \ref{eq:3zone}. We compute O/H for each object and express the results following standard convention: [O/H]+12$=\log_{10}($O/H$)+12$. We find that the \ion{H}{ii} galaxies in our sample have a median [O/H]+12 of 8.197 with an rms of 0.166 dex, while the Seyferts have a median [O/H]+12 of 8.230 with an rms of 0.249 dex. Propagating the errors in the emission-line fluxes through the calculations, we have obtained the uncertainty in the direct-method abundance for each object in our sample.  Typical uncertainties of 0.040 dex for [O/H]+12 and 0.054 dex for [N/H]+12 are found for \ion{H}{ii} galaxies.  Flux errors tend to be larger for the AGN in our SDSS sample, which results to somewhat higher typical uncertainties: 0.095 dex for [O/H]+12 and 0.132 dex for [N/H]+12.

\subsection{Comparison with the Literature}

In a complementary study, \citet{source:dors2015} compiled emission-line fluxes (including [\ion{O}{iii}] $\lambda4363$) for 47 local ($z<0.1$) Seyfert nuclei.  Based on these data, they computed electron temperatures and densities and direct-method abundances for all of the objects, which provides an opportunity to validate the abundance method outlined above.

\begin{figure}
	\includegraphics[width=\columnwidth]{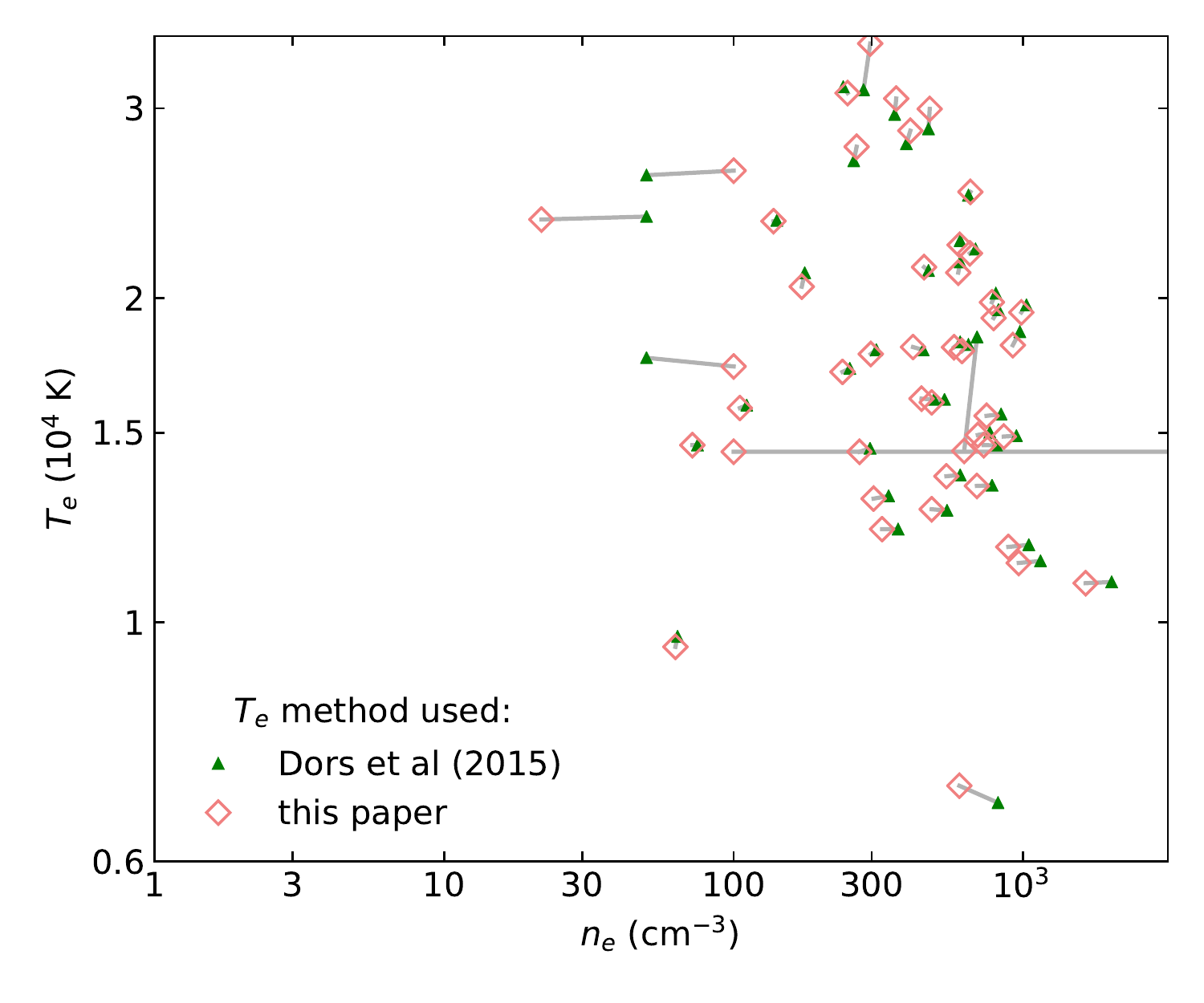}
	\caption{Comparison of electron temperatures and densities for the \citet{source:dors2015} AGN sample computed using their method (filled green triangles) and the one outlined in \S~\ref{s:phys} of this paper (open red diamonds).}
    \label{fig:dorsphyscomp}
\end{figure}

We begin by comparing the electron temperatures and densities reported for the \citet{source:dors2015} sample to those obtained for the same objects via our approach (\S~\ref{s:phys}).  This will allow us to isolate any differences in the values assumed for these quantities when we compare abundance determination methods.  As Figure \ref{fig:dorsphyscomp} demonstrates, the \citet{source:dors2015} temperatures and densities are, for the most part, in close agreement with ours. However, substantial differences exist for $\sim 20$\% of the sample:\ there are four objects for which the derived electron densities disagree by at least a factor of two, and five objects for which the electron temperatures disagree by more than $10^3$ K.  The discrepancies stem from a difference in the method used to calculate these values.  Although the \textsc{temden} routine they used computes $n_e$ iteratively (accounting for electron temperature), the method they employed to estimate $T_e$ does not simultaneously account for electron density. Moreover, the \textsc{temden} routine does not account for the difference in $T_e$ between the O$^{+2}$ and O$^{+}$ zones when calculating $n_e$.

\begin{figure}
	\includegraphics[width=\columnwidth]{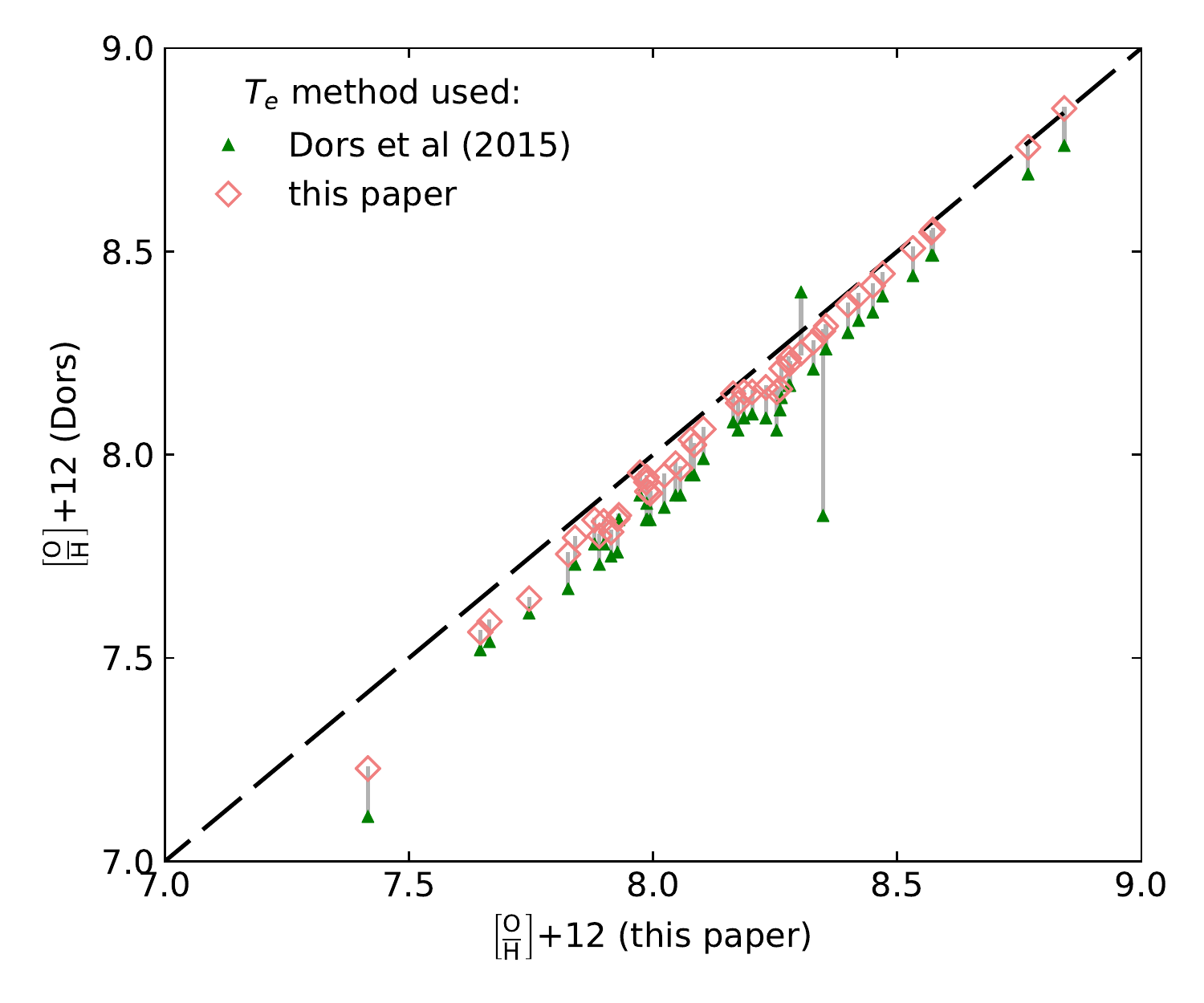}
    \caption{Direct-method oxygen abundances of AGN computed by \citet{source:dors2015} compared to those calculated for the same objects using the direct method developed in this paper. Based on the results presented in Figure \ref{fig:icf}, the abundances published by \citet{source:dors2015} have been corrected for O$^{+3}$ assuming this species represents 12.5\% of the total oxygen (filled green triangles).  Additionally, we show the values yielded by the \citet{source:dors2015} method if the approach described in \S~\ref{s:phys} for estimating electron temperatures is employed (open red diamonds).  Clearly, there is close agreement between these abundances and those derived using our method.}
    \label{fig:dorschemcomp}
\end{figure}

There are a few differences between the method for determining abundances described in \S~\ref{s:dmabn} and that presented by \citet{source:dors2015}.  The first concerns the inclusion of a correction for triply ionized oxygen.  As demonstrated in Figure \ref{fig:icf}, a considerable fraction of oxygen in AGN exists in the O$^{+3}$ state and cannot be ignored.  We therefore corrected the abundances reported by \citet{source:dors2015} assuming that 12.5\% of oxygen exists in the O$^{+3}$ state prior to comparing them to the values we obtain for their sample.  Additional differences include the approach and atomic data used to compute emissivities.  However, as Figure \ref{fig:dorschemcomp} shows, these differences are not very significant.  The median offset between the O$^{+3}$-corrected \citet{source:dors2015} abundances and those we calculate is only $\sim 0.11$ dex.  If we replace the \citet{source:dors2015} electron temperatures with those derived using the approach outlined in \S~\ref{s:phys}, the median offset drops to 0.05 dex.  We note that the discrepancy between the abundance measurements increases as [O/H]+12 decreases, which could be due to increases in electron temperature (see \S~\ref{s:cooling}).  But overall, the agreement between the \citet{source:dors2015} abundances and ours is excellent, indicating that the direct method developed here can be employed with confidence.

\section{Semi-Empirical Relations}

\subsection{Cooling Sequences}\label{s:cooling}

Metals have long been known to act as an important source of cooling in gaseous nebulae  \citep{source:spitzer1949}, and there is observational support \citep{source:kobulnicky1999} for the theoretically expected relationship between electron temperature and chemical abundance \citep[e.g.,][]{source:mcgaugh1991,source:lopezsanchez2012}. However, a comprehensive picture of the cooling process has yet to emerge.  For example, \cite{source:kobulnicky1999}  found a strong correlation between $T_e$ and metallicity using a sample of 69 extragalactic \ion{H}{ii} regions (see their Fig.~5), but did not not provide a physical description of the relationship. Conversely, \cite{source:lopezsanchez2012} derived a cooling sequence (i.e., the relation between temperature and metal coolants demonstrated by the $T_e$ vs [O/H] diagram) for model \ion{H}{ii} regions over the full range of typical metallicities, but did not compare their results to observations.  More recently,  \cite{source:nicholls2014a} have provided a mathematical description of the cooling sequence for a large sample of extragalactic \ion{H}{ii} regions and a comparison between the data and predictions of photoionization models to shed light on the physical mechanisms related to cooling. However, their results are only applicable to low-metallicity objects, i.e., those with [O/H]+12 $<8.4$. Here, we extend the characterization of cooling process to star-forming regions with [O/H]+12 $\approx 9$.

\begin{figure}
	\includegraphics[width=\columnwidth]{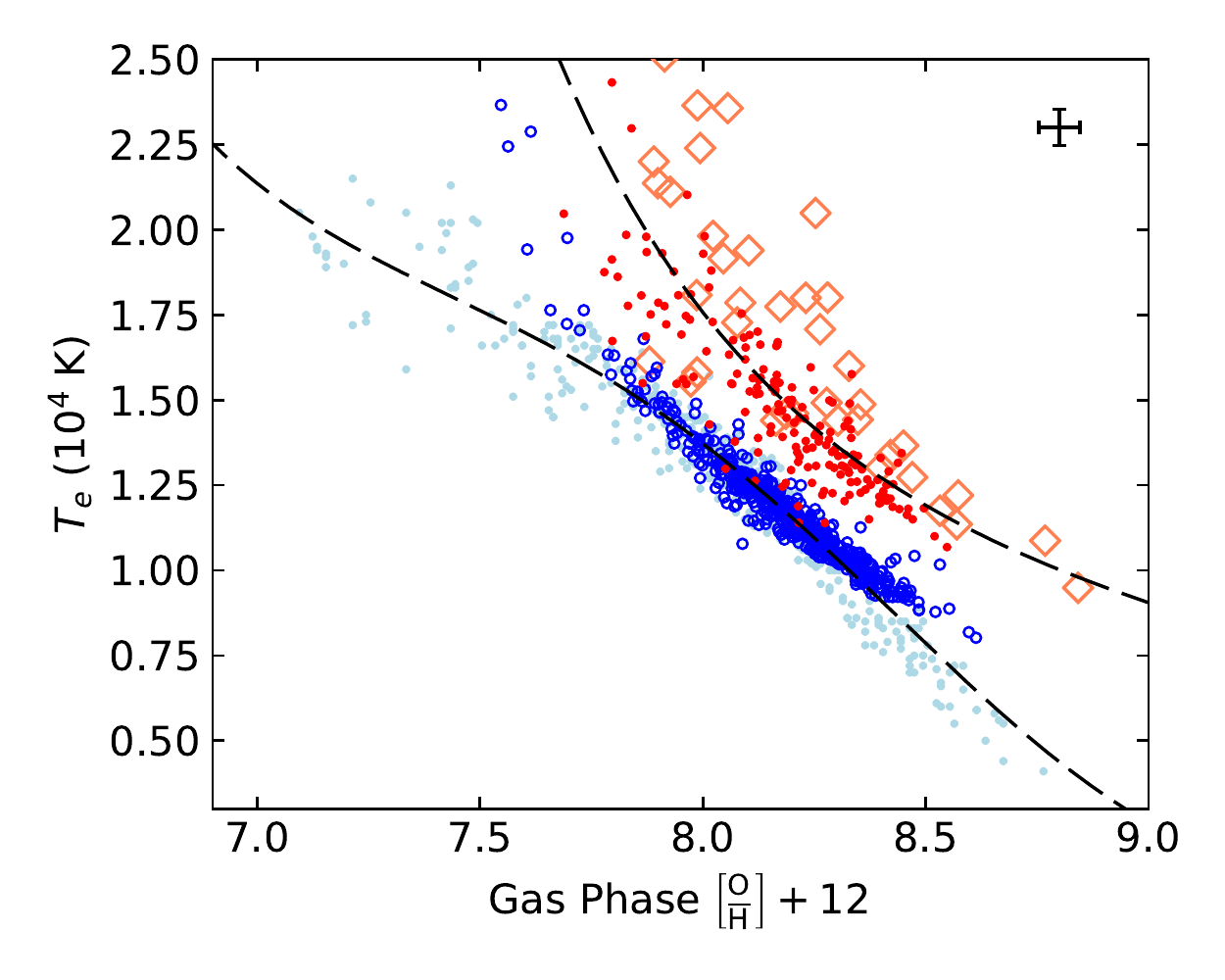}
    \caption{Electron temperature vs.\ oxygen abundance for AGN and star-forming galaxies included in several different studies. Open blue and filled red circles represent \ion{H}{ii} galaxies and AGN, respectively, from our SDSS sample with median $1-\sigma$ uncertainties shown in the upper right. Light blue circles represent extragalactic \ion{H}{ii} regions from \citet{source:pilyugin2012}, and the locations of the Dors et al.\ (2015) AGN are indicated with open red diamonds. We have corrected metallicities reported by \citet{source:pilyugin2012} and \citet{source:dors2015} to account for the O$^{+3}$ species.}
    \label{fig:oh12temp}
\end{figure}

To investigate nebular cooling over the broadest possible range of temperatures and metallicities, we have supplemented our SDSS sample with data from the literature:\ 47 Seyferts with direct-method abundances from \citet{source:dors2015} and 414 extragalactic \ion{H}{ii} regions from \cite{source:pilyugin2012}.
After correcting the \citet{source:pilyugin2012} and \citet{source:dors2015} abundances for [O$^{+3}$/H${+}$], we fitted the cooling sequences for \ion{H}{ii} regions and AGN with a third-order polynomial of the form
\begin{equation}\label{eq:cool}
y = a_3x^3 + a_2x^2 + a_1x + a_0
\end{equation}
where $x$=[O/H]$+12$ and $y=\log_{10}t_4$ (again, $t_4=T_e/ 10^4$ K). The results are displayed in Figure \ref{fig:oh12temp}, and the corresponding coefficients for each semi-empirical sequence are listed in Table \ref{tab:coolpar}. The Table also includes the coefficients obtained for fits where $x=\log_{10}t_4$ and $y=$[O/H], for the purpose of inferring [O/H] from $T_e$.

\begin{table}
	\centering
	\caption{Cooling sequence coefficients.}
	\label{tab:coolpar}
	\begin{tabular}{ccccc}
		\hline
		&\multicolumn{2}{c}{$x=$[O/H]+12} & \multicolumn{2}{c}{$x=\log_{10}t_4$}\\
		& \ion{H}{ii} & AGN  & \ion{H}{ii} & AGN\\
		\hline
		$a_0$ & ~-0.143 	& ~-0.050 		& -1.302 & -1.948\\
		$a_1$ & ~~3.171 	& ~~1.372 	& -2.917 & ~4.108\\
		$a_2$ & -23.551 	& -12.792 		& -2.500 & -4.175\\
		$a_3$ & ~58.937	& ~40.290		& ~8.304 & ~8.803\\
		\hline
	\end{tabular}
\end{table}

Figure \ref{fig:oh12temp} represents the first empirical cooling sequence for Seyfert galaxies. Ostensibly, the same cooling mechanisms are at play in both AGN and \ion{H}{ii} galaxies: collisions, recombination, and free-free interactions dissipate the kinetic energy of the electron gas in the form of electromagnetic radiation. However, the AGN sequence is characterized by higher temperatures, and it lacks the dramatic drop in $T_e$ seen in \ion{H}{ii} regions with higher metallicities. This is likely to be a consequence of the fact that AGN impart more kinetic energy per electron during the photoionization process than O and B stars do. The initially higher electron velocities reduce the rate of recombination in AGNs\footnote{As discussed in Ch. 5.11 of \citetalias{source:osterbrock2006}, recombination rates are approximately proportional to the inverse of the electron temperature, due in part to the preferential recombination of electrons with low velocities.}, which is a significant source of cooling in the O$^{+2}$ region by means of the He$^{+}+e^-\to$~He$^{0}+\gamma$ process.  Thus, the gas is prevented from dissipating the increased thermal energy \citep[\citetalias{source:osterbrock2006};][]{source:groves2004a}.

In addition to providing insight into nebular heating rates, the cooling sequences shown in Figure~\ref{fig:oh12temp} have practical applications.  When $t_4$ is treated as the independent variable, one can infer the direct-method metallicity using $T_e$ if measurements of the [\ion{O}{ii}] doublet and/or ICF helium recombination lines are not available. Similarly, if measurements of the [\ion{O}{iii}] auroral line are lacking, one could estimate the oxygen abundance by computing SEL emission coefficients for different values of [O/H] and the corresponding electron temperature indicated by the cooling sequence to recover the observed flux ratios. However, such an approach would require additional information, such as the production rate of other elements relative to oxygen.

\subsection{Nitrogen Production}\label{s:nitrogen}

Variations in \ion{H}{ii} region emission line flux ratios have been observed in spiral galaxies as early as the 1940s \citep[c.f.,][]{source:aller1942}. The \citet{source:searle1971} survey of extragalactic \ion{H}{ii} regions in the arms of four spiral galaxies demonstrated radial gradients in [\ion{O}{iii}]/\hb~and [\ion{N}{ii}]/\ha~and provided an explanation for the phenomenon: radial gradients in the oxygen and nitrogen abundances relative to hydrogen. 
Initial efforts to explore the \citet{source:searle1971} suggestion that nitrogen abundances scale with those of oxygen assumed that nitrogen is solely a secondary element \citep[e.g.][]{source:talarn1974}, meaning the predicted nitrogen abundances depend entirely on the initial stellar carbon abundances. 
\citet{source:alloin1979} introduced the first physically complete model of the co-evolution of nitrogen and oxygen. In their scheme, both primary nitrogen (i.e., dependent on carbon formed in situ by helium burning) and secondary nitrogen are significant. The \citet{source:alloin1979} predictions provided an accurate description of the [N/H]-[O/H] correlation observed at that time. Based on this framework, \citet{source:vilacostas1993} modeled a much larger sample of \ion{H}{ii} region nitrogen abundances relative to oxygen. Together with the subsequent \citet{source:vanzee1998a} study of 185 \ion{H}{ii} regions in 13 galaxies, this provided substantial evidence in support of the combined primary-secondary nitrogen production hypothesis.

\begin{figure}
	\includegraphics[width=\columnwidth]{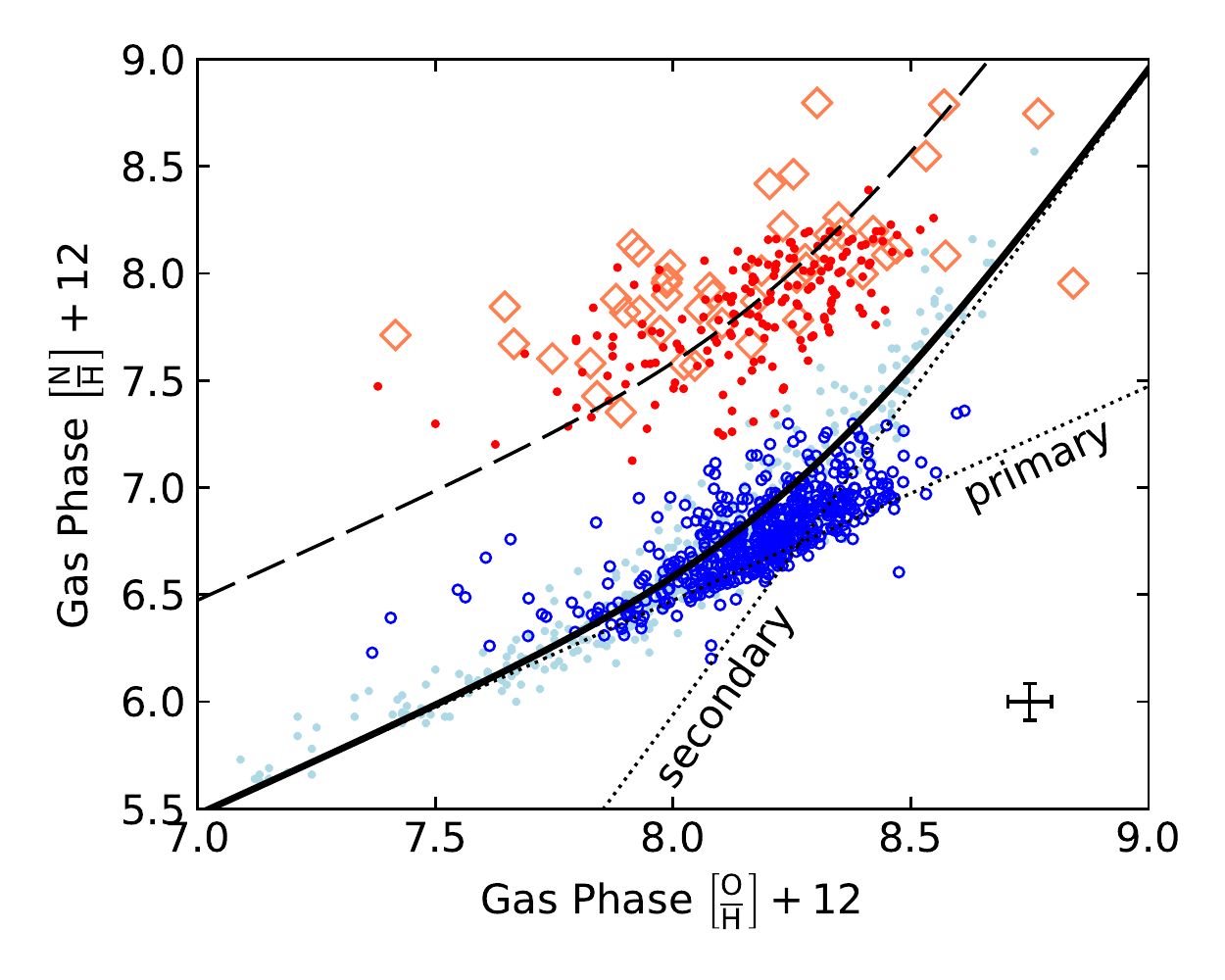}
    \caption{Nitrogen abundance vs.\ oxygen abundance for AGN and star-forming galaxies included in several different studies. Symbols as in Figure \ref{fig:oh12temp}. The solid line represents our fit to the \ion{H}{ii} nitrogen production sequence. Dotted lines represent the primary and secondary nitrogen components of the relation. The dashed lines represents the \ion{H}{ii} nitrogen production relation shifted by 0.75 dex in [N/H] and 0.125 in [O/H] to match the observed AGN abundances.}
    \label{fig:oh12nh12}
\end{figure}

To demonstrate the nitrogen-oxygen coupling for both \ion{H}{ii} galaxies and AGN, we compare the [N/H]+12 and [O/H]+12 results from both our subsample and those from the literature in Figure \ref{fig:oh12nh12}. We assume that the primary and secondary nitrogen each follow a log-linear relation \citep[e.g.,][]{source:vilacostas1993,source:groves2004a,source:gutkin2016,source:nicholls2017} and fit the \ion{H}{ii} galaxies with the sum of the two relations to obtain
\begin{equation}\label{eq:nhprod}
\frac{\text{N}}{\text{H}} = 
\frac{\text{O}}{\text{H}}\left(10^{-1.527}+10^{2[\text{O}/\text{H}]+5.940}\right).
\end{equation}
This fit is shown in Figure \ref{fig:oh12nh12} along with the individual primary and secondary nitrogen production components. 

As seen in the cooling sequences, Figure \ref{fig:oh12nh12} shows that the AGN form a unique correlation distinctly offset from that of the \ion{H}{ii} galaxies. We find the AGN nitrogen abundances are typically 1 dex higher than those in \ion{H}{ii} galaxies of the same [O/H] values. The shape of the distribution of AGN [N/H] values suggests the offset is solely in nitrogen abundance. To demonstrate this, we multiply the right side of Equation \ref{eq:nhprod} by a factor of 10 and find reasonable agreement between the adjusted description and the measured AGN abundances. 

This is not the first time nitrogen excess has been noted in AGN: \citet{source:dors2015} found a separation of as much as 2 dex between direct method [O/H] values and those inferred following \citet{source:storchiberg1998}, the latter using the nitrogen-oxygen relation of \citet{source:vilacostas1993}. To account for the discrepancy, \citet{source:dors2015} propose that this shift in [O/H] is due to temperature fluctuations or shock heating of the NLR gas, either of which would artificially inflate the estimated electron temperature and thus decrease the measured [O/H]. It is possible that infall, outflow, and star formation efficiency could collectively inflate the nitrogen abundance  \citep[e.g.,][]{source:vincenzo2016}; however, these processes only account for, at most, a few tenths of a dex in [N/O] and are likely insufficient in explaining the 1 dex difference in nitrogen abundances we measure in AGN.

While significant in demonstrating the evolution of gas phase abundances and possible complications for the direct-method abundances, the [N/H]-[O/H] diagram provides an important basis for understanding the chemical composition of nebulae. For instance, photoionization models rely heavily on the nitrogen-oxygen scaling relations \citep[e.g.,][]{source:storchiberg1998,source:groves2004a,source:gutkin2016} to make accurate predictions for flux-ratio diagnostic diagrams like the BPT because, as shown in \S~\ref{s:dmabn}, abundances directly determine emission line fluxes. If one can infer emissivity ratios from the oxygen abundance, such as we have established with the cooling sequences, then Equation \ref{eq:nhprod} provides the tool necessary for predicting [\ion{O}{iii}]/\hb~and [\ion{N}{ii}]/\ha~BPT flux ratios for a given value of [O/H], enabling the inference of metallicity solely from SELs. We explain and demonstrate this approach in detail the next section and examine the results.

\section{A New Method for Abundance Estimation}\label{s:selmethod}

Previous approaches to determining relative abundances have relied on either SEL-abundance correlations for \ion{H}{ii} regions and galaxies \citep[e.g.,][]{source:vanzee1998b,source:pp04}, which have not been validated for AGN, or radiative transfer models, which require many assumptions about the ionizing source and the physical conditions present in the nebula \citep[e.g.,][]{source:storchiberg1998,source:dors2017}. In this section, we present a new method for estimating direct-method oxygen abundances for AGN based solely on the SEL flux ratios that are routinely employed in BPT diagnostic diagrams.

\subsection{A Simple Equilibrium Model}\label{s:selmodel}

Our SEL approach for estimating abundances involves a reverse-engineering of the direct method outlined above in \S~\ref{s:abn}. We begin by using Equations \ref{eq:colemit} and \ref{eq:recemit} to obtain (a) the [\ion{N}{ii}]/\ha\ flux ratio
\begin{equation}\label{eq:n2hasimp}
\frac{j_{\lambda6583}}{j_{\lambda6563}} = 
\frac{{\text{N}^{+}}}{{\text{H}^+}}
\frac{\varepsilon_{\lambda6583}(T_e,n_e)}{\alpha_{\lambda6563}(T_e)}
\end{equation}
from the relative ionic abundance N$^+$/H$^+$ and the [\ion{O}{ii}] electron temperature (see \S~\ref{s:tempdens}), and (b) the [\ion{O}{iii}]/\hb\ flux ratio
\begin{equation}\label{eq:o3hbsimp}
\frac{j_{\lambda5007}}{j_{\lambda4861}} = 
\frac{\text{O}^{+2}}{\text{H}^+}
\frac{\varepsilon_{\lambda5007}(T_e,n_e)}{\alpha_{\lambda4861}(T_e)}
\end{equation}
from the relative ionic abundance O$^{+2}$/H$^+$ and the [\ion{O}{iii}] electron temperature.  The process of estimating the emissivity ratios in these two equations is simplified by computing each value of $\varepsilon$ and $\alpha_{eff}$ with \textsc{PyNeb} assuming a typical electron density of 100 cm$^{-3}$ (see \S~\ref{s:phys}). The ratio of emissivities can then be fitted with the analytic expression
\begin{equation}\label{eq:emissratio}
\log_{10}\left(\frac{\varepsilon_{\chi^i}}{\varepsilon_{\text{H}^+}}\right)
= \frac{a_\varepsilon}{t_4+b_\varepsilon}+c_\varepsilon,
\end{equation}
to obtain a relation between emissivity and electron temperature.  The coefficients associated with the [\ion{N}{ii}]/\ha\ and [\ion{O}{iii}]/\hb\ flux ratios are listed in Table~\ref{tab:emissparams}.

\begin{table}
	\centering
	\caption{BPT emissivity ratio parameters for Equation \ref{eq:emissratio}.}
	\label{tab:emissparams}
	\begin{tabular}{lccr} 
		\hline
		$\log_{10}({\varepsilon_{\chi^i}}/{\varepsilon_{\text{H}^+}})$ & $a_\varepsilon$ & $b_\varepsilon$ & $c_\varepsilon$ \\
		\hline
		$[$\ion{N}{ii}$]$/\ha & -1.003 & 9.800$\times10^{-2}$ & 5.195 \\
		$[$\ion{O}{iii}$]$/\hb & -1.381 & 6.845$\times10^{-5}$ & 5.845 \\
		\hline
	\end{tabular}
\end{table}

As we have related electron temperature and metallicity by means of the cooling sequences (\S~\ref{s:cooling}), we can eliminate the temperature dependence of Equation \ref{eq:emissratio} by substituting Equation \ref{eq:cool} (with the parameters for $x=$[O/H]+12 listed in Table~\ref{tab:coolpar}) for $t_4$. This allows the emissivity ratios for [\ion{N}{ii}]/\ha\ and [\ion{O}{iii}]/\hb\ to be computed for a given value of [O/H]+12. Furthermore, the [N/H]-[O/H] coupling established in \S~\ref{s:nitrogen} allows us to express the nitrogen abundance as a function of the oxygen abundance via Equation \ref{eq:nhprod}. All that remains to be determined is the fraction of the total abundance associated with each ionic species.

The ionic abundances can be expressed in terms of total abundances to indicate the fraction of a given metal in a particular state of ionization, i.e., the \textit{ion fraction}. For doubly-ionized oxygen, we define the ion fraction $\eta$ such that
\begin{equation}\label{eq:ionfrac}
\eta=\frac{\text{O}^{+2}}{\text{O}},
\end{equation}
which gives
\begin{equation}\label{eq:highionfrac}
\frac{\text{O}^{+2}}{\text{H}^+} = \eta \frac{\text{O}}{\text{H}}.
\end{equation}
The ion fraction for N$^+$ can be expressed in terms of $\eta$ and an addition variable $\zeta$, which accounts for the fraction of oxygen contianed in the O$^{+3}$ species, such that, ultimately,
\begin{equation}\label{eq:lowionfrac}
\frac{\text{N}^{+}}{\text{H}^+} =  \left(1-\eta - \zeta\right)\frac{\text{N}}{\text{H}}.
\end{equation}
The ICF computed from helium recombination lines (see Figure \ref{fig:icf}) establishes the characteristic fraction of oxygen contained in the O$^{+3}$ species for AGN and \ion{H}{ii} galaxies:
\begin{equation}
\zeta \approx
\left\{
\begin{array}{ll}
0.01~	&\text{\ion{H}{ii}}\\
0.125	&\text{AGN}\\
\end{array}
\right. .
\end{equation}
As $\eta$ relates total abundances to ionic abundances, values of the [\ion{N}{ii}]/H$\alpha$ (Eq.\ \ref{eq:n2hasimp}) and [\ion{O}{iii}]/H$\beta$ (Eq.\ \ref{eq:o3hbsimp}) flux ratios can be expressed solely in terms of [O/H] and $\eta$ for each type of object:
\begin{equation}\label{eq:n2hafull}
\frac{j_{\lambda6583}}{j_{\lambda6563}} = 
\left(1-\eta-\zeta\right)\frac{\text{N}}{\text{H}}
\left(\frac{\text{O}}{\text{H}}\right)
\frac{\varepsilon_{\lambda6583}}
{\alpha_{\lambda6563}}(t_4(\text{O}/\text{H}))
\end{equation}
and
\begin{equation}\label{eq:o3hbfull}
\frac{j_{\lambda5007}}{j_{\lambda4861}} = 
\eta\frac{\text{O}}{\text{H}}
\frac{\varepsilon_{\lambda5007}}
{\alpha_{\lambda4861}}(t_4(\text{O}/\text{H}))
\end{equation}

\begin{figure}
	\includegraphics[width=\columnwidth]{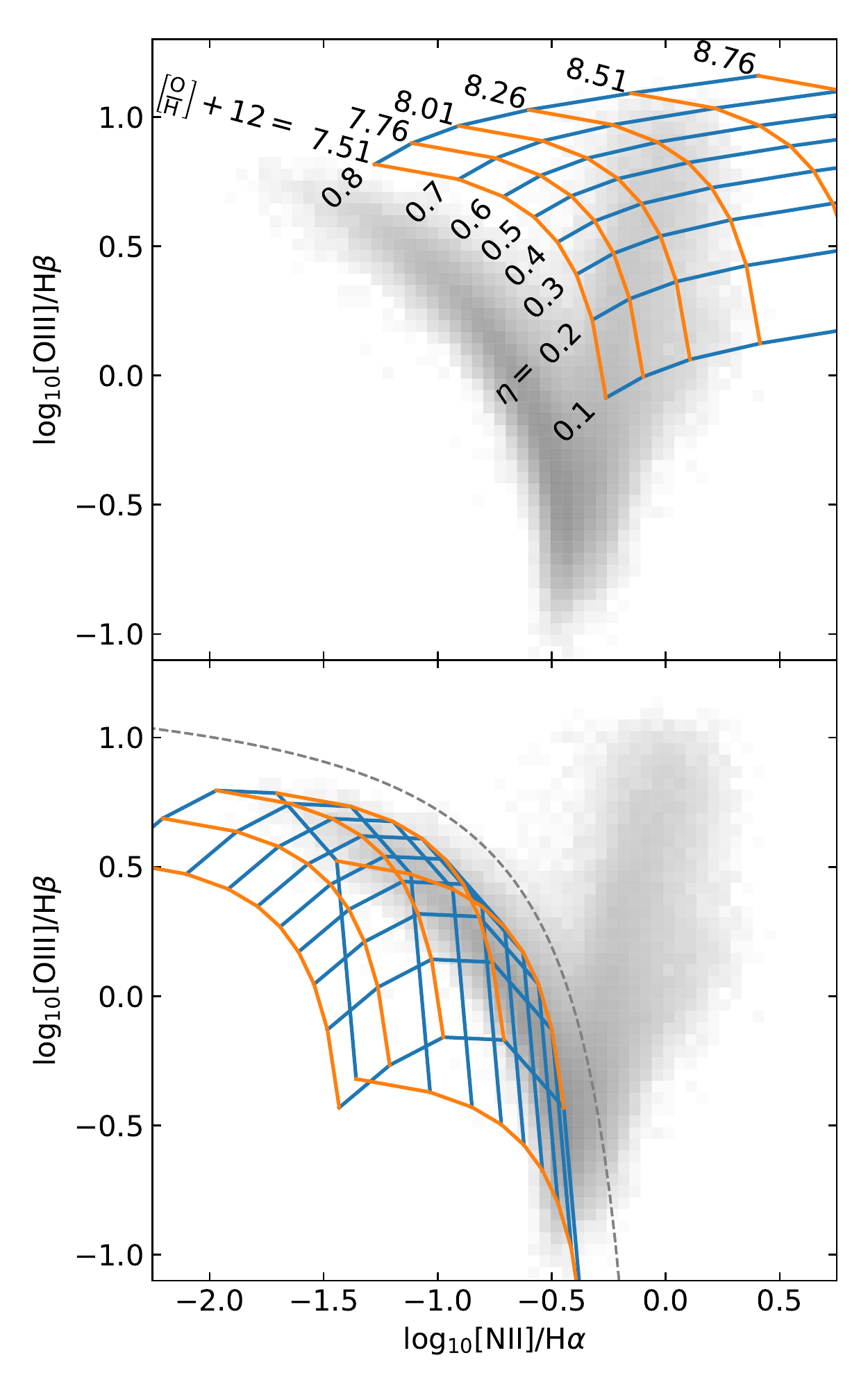}
    \caption{Grids of flux ratio predictions based on our simple equilibrium model for AGN (top) and \ion{H}{ii} regions (bottom).  Shown in grey are line ratios for the objects in our parent SDSS sample. Values of [O/H]+12 range from 7.51 to 8.76 in increments of 0.25 dex. Values of $\eta$, the O$^{+2}$ ion fraction, range from 0.1 to 0.9 for \ion{H}{ii} galaxies and from 0.1 to 0.8 for AGN.  The short-dashed curve in the lower panel indicates the empirical demarcation between AGN and star-forming galaxies from \citet{source:kauffmann2003}.}
    \label{fig:simpgrids}
\end{figure}

Using these equations, we have calculated grids of predicted flux ratios for \ion{H}{ii} galaxies and AGN over reasonable ranges of ion fraction and oxygen gas phase abundance.  These grids are overlaid on the [\ion{O}{iii}]/H$\beta$ vs.\ [\ion{N}{ii}]/H$\alpha$ diagnostic diagram in Figure \ref{fig:simpgrids}.  It is clear from the figure that our predictions are consistent with the flux ratios observed for both populations.

A close examination of the \ion{H}{ii} galaxy predictions shown in Figure \ref{fig:simpgrids} reveals that, for a fixed value of $\eta$, the [\ion{O}{iii}]/\hb\ flux ratio increases with [O/H]+12 at first, but then declines precipitously at higher metallicities \citep[cf. the photoionization model predictions of constant ionization parameter in ][]{source:stasinska2006}.  The maximum in [\ion{O}{iii}]/\hb\ occurs at [O/H]+12 = 8.124 for all values of $\eta$, which, according to the \ion{H}{ii} cooling sequence (see Eq.\ \ref{eq:cool} and Table~3) corresponds to a threshold electron temperature of $T_{e,{\rm thresh}} = 1.2 \times 10^4$ K.  Above this metallicity, the increase in oxygen abundance is offset by a decrease in electron temperature, which results in fewer collisionally excited O$^{+2}$ ions overall and, thus, a decrease in the [\ion{O}{iii}]/\hb\ ratio.  This behavior causes the grid of predicted line ratios to fold over on itself, in the sense that line ratios for objects with extremely high values of $\eta$ and [O/H]+12 would nearly overlap with those for which both $\eta$ and [O/H]+12 are very low.

The apparent ``crease'' in the fold roughly follows the locus of observed points for \ion{H}{ii} galaxies, along which clear trends in $\eta$ and [O/H]+12 are evident.  For example, as log [\ion{O}{iii}]/\hb\ increases from $-0.75$ to 0.0 to $+0.75$, $\eta$ increases from $\lesssim 0.3$ to $\sim 0.2 - 0.6$ to $\gtrsim 0.75$, and [O/H]+12 decreases from $\gtrsim 8.5$ to $\sim 8.5$ to $\lesssim 8.25$.  Following predictions from photoionizatoin models, others have proposed that metallicity is a direct cause for the location and shape of the \ion{H}{ii} sequence \citep[e.g.,][]{source:stasinska2006,source:kewley2013}, which is consistent with the trend we see in [O/H]+12 in our grid of flux ratios.  Assuming that OB stars in \ion{H}{ii} galaxies and the nebulae they power have similar metallicities, the run of $\eta$ with [O/H]+12 along the star-forming locus can be understood as a consequence of stars' ionizing radiation decreasing with an increase in metallicity by means of physical processes such as line blanketing, which is consistent with the current understanding of stellar ultraviolet continua \citep[e.g.,][]{source:levesque2010,source:kewley2013}.

We note that there is scatter in the data used to derive the cooling sequence (Eq.~\ref{eq:cool}) and nitrogen production relation (Eq.\ \ref{eq:nhprod}).  Accounting for this scatter would permit a maximum shift in the \ion{H}{ii} line-ratio grid of about +0.1 dex in [\ion{N}{ii}]/\ha\ and +0.05 dex in [\ion{O}{iii}]/\hb . This shift would place the ``crease'' in the grid along, but not over, the empirical curve from \cite{source:kauffmann2003} that separates \ion{H}{ii} galaxies and other types of emission-line objects on the BPT diagram.  Moreover, the crease in our grid closely matches the upper limit predicted by \citet{source:stasinska2006} using photoionization models. This close proximity of our model flux ratios to empirical and theoretical predictions bolsters our confidence in the accuracy of the flux ratio predictions. Given the dependence of the grid's location on nitrogen abundance and electron temperature, we surmise that the breadth of the \ion{H}{ii} galaxy locus on the diagram is due in part to small variations in [N/H] and $T_e$ amongst objects with the same value of [O/H].

As Figure \ref{fig:simpgrids} indicates, the region of the BPT diagram occupied solely by AGN is largely covered by the grid of predicted flux ratios.  Moreover, at the lowest metallicities considered, our predictions for AGN fall above the \cite{source:kauffmann2003} curve. Note that for a fixed value of $\eta$, the [\ion{O}{iii}]/\hb\ ratio does not turn over as [O/H]+12 increases, so there is no ``folding'' of the model grid as in the case for \ion{H}{ii} galaxies.  This is due to the fact that $T_{e,{\rm thresh}}$ occurs at a substantially higher metallicity in the AGN cooling sequence.  The decrease in electron temperature with increasing [O/H] results instead in the modest gradient seen in [\ion{O}{iii}]/\hb.  As discussed below, the lack of folding of the AGN grid is quite fortunate:\ it ensures that pairs of line ratios map uniquely to values of the oxygen abundance.

\subsection{Comparison to the Direct Method}

\begin{figure}
	\includegraphics[width=\columnwidth]{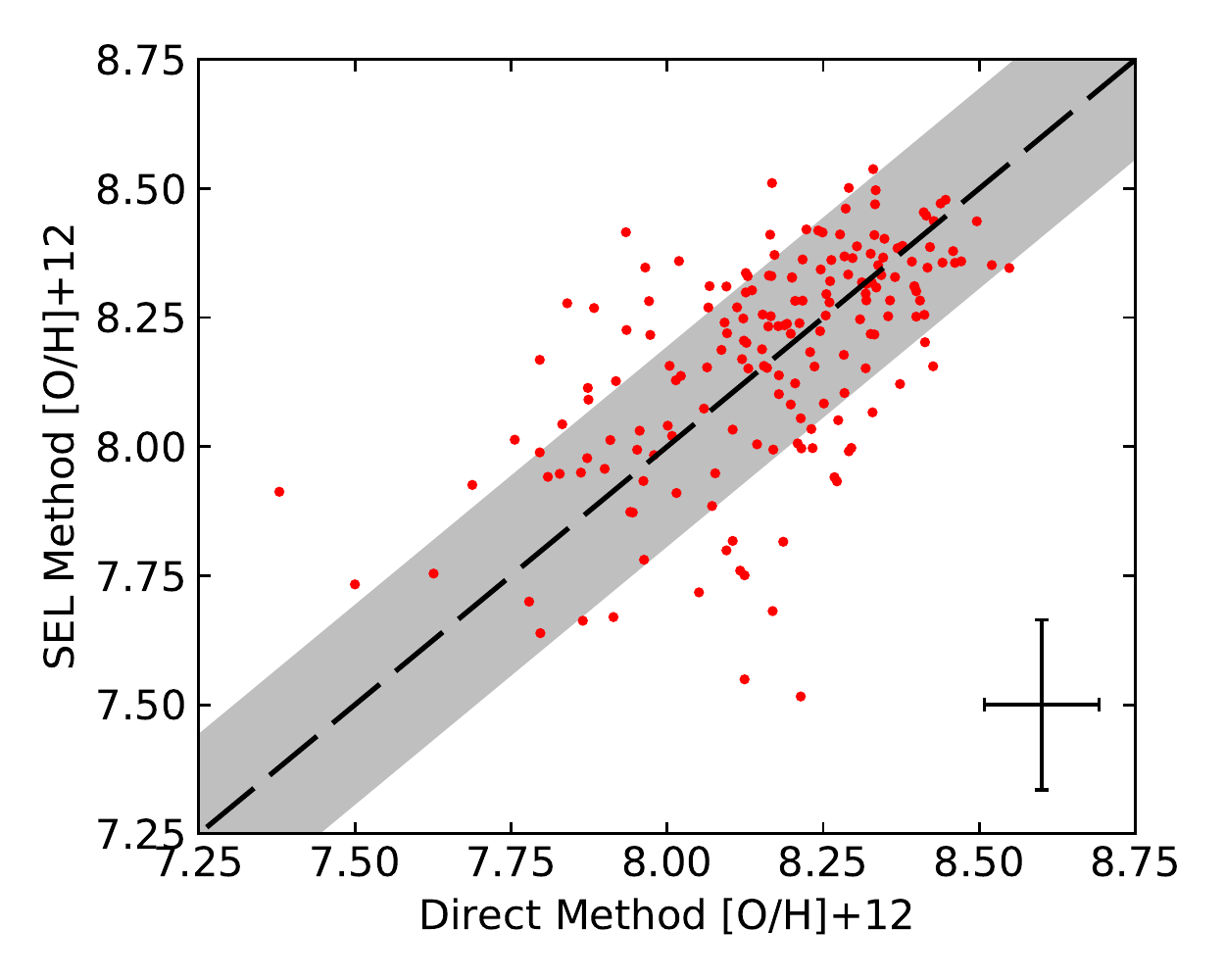}
    \caption{Comparison of [O/H]+12 values estimated from our SEL approach to those obtained via the direct method for the SDSS AGN sample.  The width of the grey region, which includes 68\% of the sample, indicates that application of the SEL method adds an uncertainty of 0.15 dex to the error in the direct method abundances of AGNs.  The total 1~$\sigma$ error in the SEL abundances has a typical value of 0.18 dex.}
    \label{fig:oh12compare}
\end{figure}

To obtain AGN metallicities from our SEL model, we employ Equations \ref{eq:n2hafull} and \ref{eq:o3hbfull} to generate a grid an order of magnitude finer than that shown in Figure \ref{fig:simpgrids} and compute the two-dimensional bicubic interpolation of [O/H]+12 (and, separately, of $\eta$) with respect to the predicted flux ratios.  For our sample of SDSS AGN, we have compared abundances estimated this way to those determined via the direct method described in \S\S~\ref{s:dmabn}--\ref{s:icf}.  The results, displayed in Figure \ref{fig:oh12compare}, indicate that the agreement between SEL and direct method abundances is very good.  The horizontal error bar in the Figure represents the median uncertainty in the direct-method values of [O/H] (\S~\ref{s:icf}).  The width of the grey region on the diagram, which includes 68\% of the points, implies that on average an additional error of 0.15 dex (shown by the vertical error bar) is incurred when the SEL method is applied.  In total, a typical 1~$\sigma$ uncertainty of $\sim 0.18$ dex is expected for the SEL abundances.  While this may be insufficient for certain applications, our SEL method provides a means to obtain a robust, empirically based metallicity estimate for any AGN classified via the [\ion{O}{iii}]/\hb\ vs.\ [\ion{N}{ii}]/\ha\ BPT diagram, which previously has not been possible.

\subsection{Extension to the Parent Sample}

\begin{figure}
	\includegraphics[width=\columnwidth]{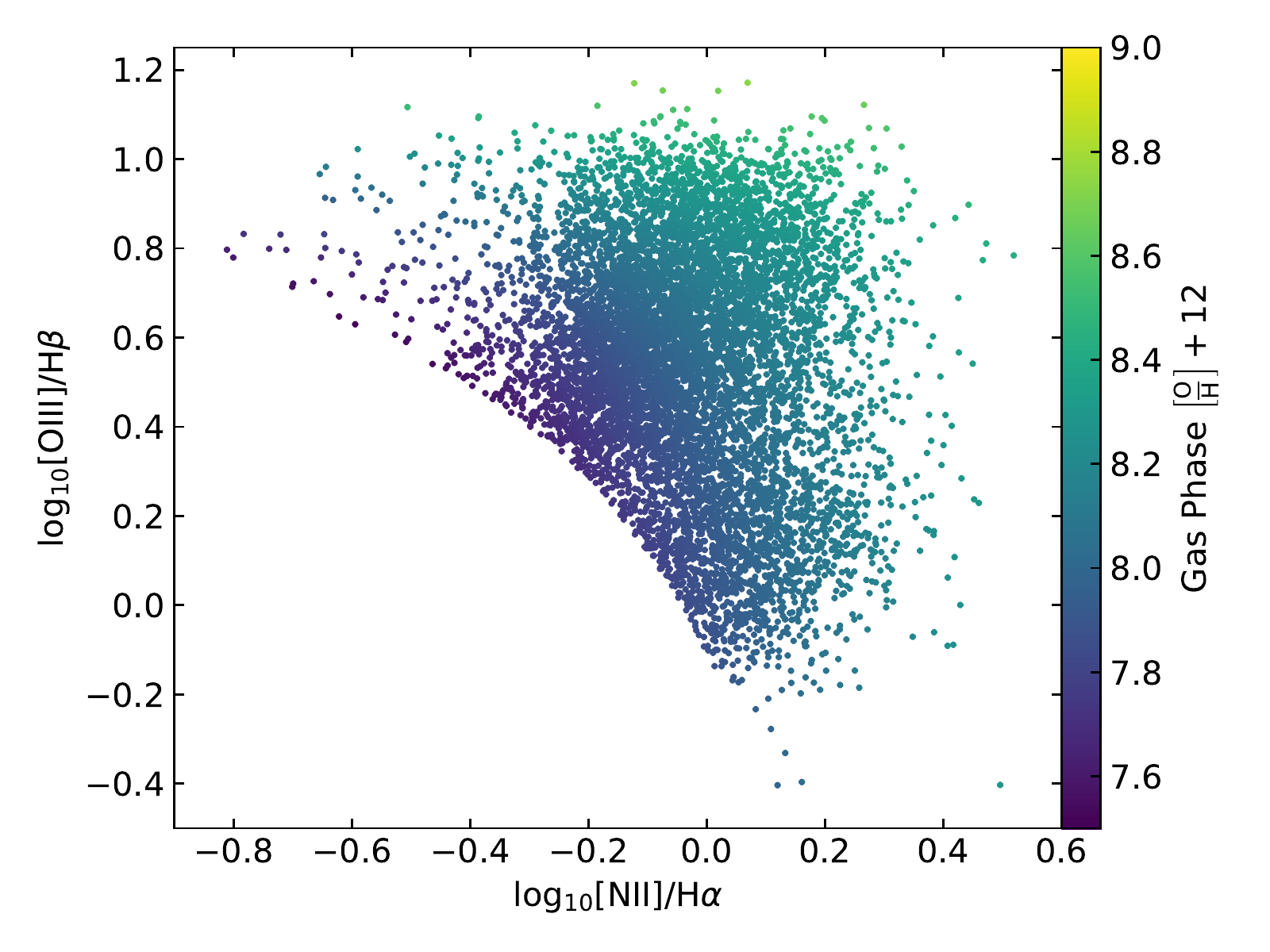}
    \caption{Oxygen abundances of the 8685 AGN in our SDSS parent sample, plotted according to their locations on the BPT diagnostic diagram.}
    \label{fig:bptoh12}
\end{figure}

To illustrate the general utility of our SEL abundance method, we have applied it to the 8,720 AGN in the SDSS parent sample described in \S~\ref{s:parentsample}.  Color-coded metallicities for these objects are plotted as a function of their BPT flux ratios in Figure \ref{fig:bptoh12}.  As anticipated from Figure \ref{fig:simpgrids}, the values of [O/H]+12 observed for AGN span more than an order of magnitude.  Low-metallicity AGN (with predominantly high values of $\eta$) are present, but are rare.  This is due in part to the quasi-flux limited nature of the parent sample and fact that the low-metallicity AGN tend to have low emission-line luminosities.  The region in the diagram where metallicity is high and $\eta$ is low is devoid of objects.

In order to simplify the process of calculating AGN oxygen abundances for pairs of [\ion{O}{iii}]/\hb\ and [\ion{N}{ii}]/\ha\ ratios, we used singular value decomposition to fit the grid of flux-ratio predictions used to construct Figure \ref{fig:bptoh12} with a two-dimensional polynomial:
\begin{eqnarray}\label{eq:oh12bpt2d}
z =
&7.863 +1.170x+0.027y-0.369x^2\nonumber\\
&+0.208y^2-0.406xy-0.100x^3\\
&+0.323y^3+0.354x^2y-0.333xy^2\nonumber
\end{eqnarray}
where $x = \log$ [\ion{N}{ii}]/\ha, $y = \log$ [\ion{O}{iii}]/\hb, and $z =$ [O/H]+12.  In the [O/H]+12 $= 7.5 - 9.0$ range, abundances calculated from the polynomial fit are nearly indistinguishable from those derived via interpolation of the grid of predicted line ratios.

\subsection{Implications for LINERs}

The class of emission-line galaxies known as LINERs \citep{source:heckman1980} occupy the lower portion of the AGN region on the [\ion{O}{iii}]/\hb\ vs.\ [\ion{N}{ii}]/\ha\ BPT diagram.  As \citet{source:molina2018} discuss, a number of different processes can, and probably do, give rise to LINER-like emission lines in galaxies.  Strictly speaking, the AGN abundances predicted from our SEL method are valid only for objects that are powered by photoionization by a hard, non-stellar continuum associated with accretion onto a supermassive black hole.  Unfortunately, it is generally not possible to identify the primary power source in LINERs from their optical spectra alone.  Therefore, some caution is warranted when employing Equation~\ref{eq:oh12bpt2d} to estimate [O/H]+12 for objects that reside in the LINER region of the BPT diagram.

\begin{figure}
	\includegraphics[width=\columnwidth]{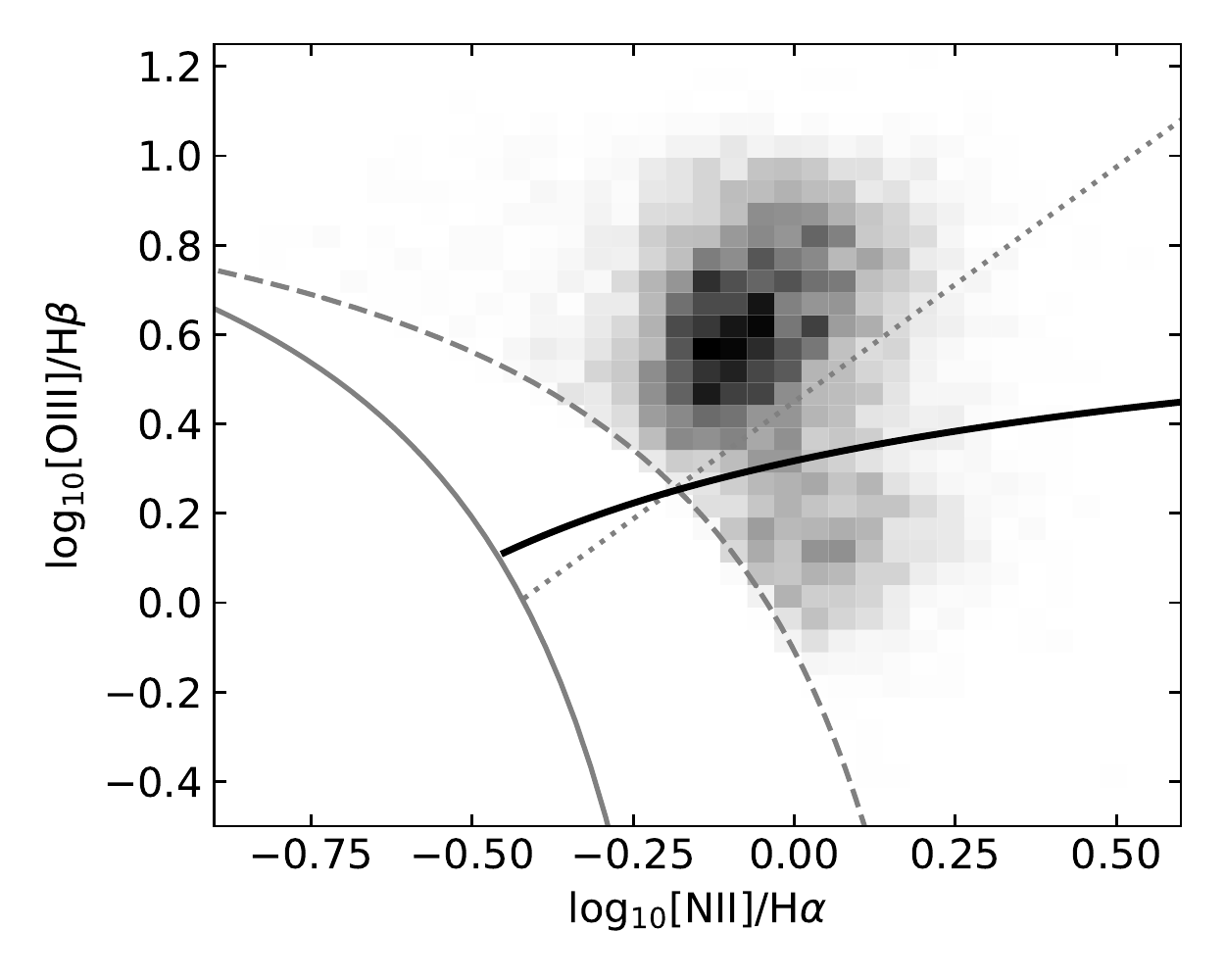}
    \caption{Distributions of AGN flux ratios for the SDSS parent sample. The darkest areas indicate where the densities of objects are greatest.  Empirical \citep{source:kauffmann2003} and theoretical \citep{source:kewley2001} demarcations between \ion{H}{ii} galaxies and AGN are shown with solid and dashed grey curves, respectively. In the AGN zone, the empirical division between Seyfert galaxies and LINERs proposed by \citet{source:schawinski2007} is indicated with a dotted line.  The solid black curve, which represents the predictions of our SEL model for $\eta=0.190$, appears to be more effective at discriminating between the two classes of objects.}
    \label{fig:bptdemarc}
\end{figure}

It has, however, been argued that the majority of LINERs are in fact bona fide AGN \citep{source:ho2008}.  If this is true of LINERs in the SDSS parent sample, the SEL model outlined in \S~\ref{s:selmodel} lends insight into how these objects and classical Seyfert nuclei might be distinguished.  Interpolating $\eta$ over the flux ratios predicted for AGN from Equations \ref{eq:n2hafull} and \ref{eq:o3hbfull} (as we did above for [O/H]+12), we find that the distribution of $\eta$ values for the parent sample is bimodal, with peaks at $\eta=0.107$ and $\eta=0.312$ and a deep minimum at $\eta=0.190$. This bimodality in $\eta$ suggests that AGN in the parent sample belong to one of two distinct ionization groups:\ Seyferts ($\eta>0.190$) or LINERs ($\eta<0.190$).  The two-dimensional histogram of line ratios shown in Figure \ref{fig:bptdemarc} confirms that the parent sample AGN are concentrated into two well-separated groups on the BPT diagram.  Furthermore, it is clear from the Figure that the curve associated with the SEL model for $\eta = 0.190$ provides a viable discriminator for Seyferts and LINERs.  The curve can be approximated by the expression
\begin{equation}\label{eq:demarc}
y = \frac{-0.628}{x+1.417}+0.753
\end{equation}
where $x$ and $y$ are the base-ten logarithms of the [\ion{N}{ii}]/\ha~and [\ion{O}{iii}]/\hb~flux ratios, respectively.  Note that at the intersection with the \citet{source:kewley2001} ``maximum starburst'' line, the $\eta = 0.190$ curve overlaps with the empirical Seyfert/LINER boundary proposed by \citet{source:schawinski2007}.  But overall, the $\eta = 0.19$ curve appears to delineate the two classes of objects better, and it has the advantage of being physically tied to photoionization through the O$^{+2}$ ion fraction, which gives rise to the [\ion{O}{iii}] $\lambda$5007 emission line.

\section{Conclusion}\label{s:future}

Metallicity estimates for AGN are necessary for a complete picture of the properties and evolution of the nuclear environments of galaxies. In this paper, we have derived measurements for the physical and chemical properties of SDSS AGN based on a simple three-zone ionization structure. From the inferred oxygen abundances and corresponding electron temperatures, we have obtained the first empirical cooling sequence and nitrogen production relations for Seyfert galaxies. Using the empirically-calibrated relations revealed by these results, we have reframed traditional emission-line flux-ratio diagnostic diagrams as probes of temperature and chemical composition in addition to ionization structure. Based on this insight, we have developed a method for estimating the direct-method abundances of AGN solely from the strong emission lines (SELs) commonly employed in BPT diagnostic diagrams. Our method can be expressed by a single equation as a function of observed SEL flux ratios, which is effective at recovering the direct-method abundances within the uncertainties of the measurements. We have applied our new method to $\sim8700$ AGN with detected SELs in the SDSS DR8, which represents the largest sample of AGN with direct-method abundances to date. The use of SEL line ratios readily facilitates investigations into the evolution of chemical abundances in AGN.
Although not our primary objective, our model also provides a viable, physically-motivated discriminator between Seyferts and LINERs. 

Future work which might lead to refinements in this method include:
\begin{enumerate}
    \item Follow-up spectroscopy of AGN for which measurements of the [\ion{O}{ii}] doublet are lacking in order to extend the empirical relations to the local volume;
    \item Estimates of the low-ionization temperature from the [\ion{N}{ii}] or [\ion{O}{ii}] auroral lines to obviate the electron temperature scaling employed in \S\ref{s:etemp}, followed by estimates of thermal inhomogeneities from computed temperatures to improve direct-method abundances;
    \item More detailed comparisons between the results of our method and those found with object-specific photoionization models, general methods such as the \citet{source:storchiberg1998} and \citet{source:perezmontero2019} approaches, and SEL inference methods such as the traditional $R_{23}$ diagnostic;
    \item New photoionization modeling to confirm that the three-zone model is sufficient in recovering the total oxygen abundance.
\end{enumerate}
While the results of this study prompt additional investigation, we have laid an important foundation for inferring the chemical abundances and ion fraction in AGN. Moreover, the cooling sequence and nitrogen abundances in AGN have opened new opportunities for understanding chemical evolution and cooling processes in the ISM of galactic nuclei. Ultimately, chemical abundance estimates for large populations of AGN, combined with stellar population synthesis, will improve our understanding of the relationship between metallicity and star formation in AGN environments\textemdash insight that is essential for investigations of the AGN-starburst connection and the coevolution of AGN and their host galaxies.





\bibliographystyle{mnras}
\bibliography{bib}




\appendix

\section{Reddening Corrections}\label{a:dered}

In the Portsmouth reduction pipeline, emission-line fluxes are corrected for extinction by applying the reddening inferred from their ssp template fit to the starlight continuum.  In the majority of cases, however, this fails to account fully for the attenuation of the emission lines. Emission-line extinction corrections typically involve a comparison between observed and intrinsic values of the Balmer decrement $F_{\text{H}\alpha}/F_{\text{H}\beta}$. For an assumed reddening law $f(\lambda)$, the equation of radiative transfer becomes
\begin{equation}\label{eq:red}
\frac{F_{\text{H}\alpha}}{F_{\text{H}\beta}} = 
\frac{F_{\text{H}\alpha,0}}{F_{\text{H}\beta,0}} 10^{-c\left[f(\text{H}\alpha)-f(\text{H}\beta)\right]}
\end{equation}
where the subscript 0 denotes the intrinsic Balmer line fluxes and $c$ is a scale factor related to the amount of extinction suffered by the observed radiation \citepalias[][]{source:osterbrock2006}. However, the intrinsic Balmer decrement depends (albeit weakly) on electron temperature, so it is necessary to compute or assume an initial estimate of $T_e$ in order to solve for $c$. 

Ideally, the process should be iterative, following these steps: ($i$) estimate the electron temperature, ($ii$) determine the expected value of the intrinsic Balmer decrement given this estimate of $T_e$, ($iii$) solve for $c$ (from which the color excess $E[B-V]$ can be derived), ($iv$) correct the emission-line fluxes for extinction and recompute $T_e$. We have explored this iterative technique by performing a Monte Carlo simulation based on $10^4$ randomly generated observed Balmer decrements between 2.75 and 6, electron temperatures between 3,000 K and 30,000 K, and electron densities between 1 and 1,000 cm$^{-3}$. Values of the intrinsic Balmer decrement were computed using \textsc{PyNeb} to calculate the Case B emissivities for H$\alpha$ and H$\beta$. For estimating $E(B-V)$, we adopted the \cite{source:calzetti2001} extinction law.

The results of our simulation reveal a strong correlation between $E(B-V)$ and the observed Balmer decrement, which can be described as 
\begin{equation}\label{eq:ebmvha}
E(B-V) = 1.943\log_{10}\left(\frac{F_{\text{H}\alpha}}{F_{\text{H}\beta}}\right)
				-\left(0.81+\frac{0.1}{0.266+t_4}\right),
\end{equation}
where $t_4 = T_e/10^4$~K. This expression is not strongly dependent on electron temperature, but overall its application yields better estimates of the intrinsic emission-line fluxes than those we would obtain assuming a single fixed value for the electron temperature.


\bsp	
\label{lastpage}
\end{document}